\documentclass[sigplan,nonacm]{acmart}

\def\arxivsubmission{}

\AtBeginDocument{
  }

\usepackage{enumitem}
\usepackage{pifont}
\usepackage{makecell}
\usepackage{hyperref}
 \usepackage[export]{adjustbox}
\usepackage{xspace}
\usepackage{subcaption}
\usepackage{amsmath}
\usepackage{amsthm}
\usepackage[linesnumbered, noend]{algorithm2e}
\usepackage{algpseudocode}
\usepackage{multirow}

\usepackage{listings}
\ifdefined\arxivsubmission
  \usepackage[frozencache,cachedir=.]{minted}
\else
  \usepackage{minted}
\fi
\usepackage[skins]{tcolorbox}
\usepackage{xcolor}

\def\ie{\textit{i.e.},\xspace}
\def\eg{\textit{e.g.},\xspace}
\def\etc{\textit{etc.}\xspace}

\newcommand{\systemName}{\texttt{FMplex}\xspace}
\newcommand{\vFM}{\texttt{vFM}\xspace}
\newcommand{\vFMs}{\texttt{vFMs}\xspace}

\newcommand{\FMAPI}{\texttt{Task-API}\xspace}
\newcommand{\systemEngine}{\texttt{FMplex-Controller}\xspace}

\newcommand{\SingleTask}{\texttt{ST}\xspace}
\newcommand{\BestEffort}{\texttt{BE}\xspace}
\newcommand{\SpatialPartition}{\texttt{SP}\xspace}

\newcommand{\SharedBE}{\texttt{S-BE}\xspace}

\newcommand{\SharedSTFQ}{\texttt{S-STFQ}\xspace}

\usepackage{tikz}
\newcommand*\circled[1]{\tikz[baseline=(rc.base)]{
  \node[shape=circle, fill=black, draw=black, line width=0.3pt,
        inner sep=0.6pt, text height=1.0ex, text depth=0.15ex] (rc)
    {\scriptsize\color{white}#1};}}

\definecolor{lightblue}{HTML}{90D5FF}
\newcommand*\reqcircle[1]{\tikz[baseline=(rc.base)]{
  \node[shape=circle, fill=lightblue, draw=black, line width=0.3pt,
        inner sep=0.6pt, text height=1.0ex, text depth=0.15ex] (rc)
    {\scriptsize\color{black}#1};}}
\newtcolorbox{keytakeawaybox}{
  enhanced,
  colback=green!10,
  colframe=green!10,
  boxrule=0pt,
  arc=3pt,
  left=0pt, right=0pt, top=0pt, bottom=0pt,
  fontupper=\itshape
}
\newcommand{\keytakeaway}[1]{
  \begin{keytakeawaybox}\textbf{Key Takeaway.} #1\end{keytakeawaybox}
}
\DeclareTextFontCommand{\texttt}{\ttfamily\upshape}

\renewcommand\footnotetextcopyrightpermission[1]{}
\acmISBN{978-1-4503-XXXX-X/2018/06}

\begin{document}

\title{
\systemName: Model Virtualization for Serving Extensible Foundation Models
}

\author{
    Hetvi Shastri\textsuperscript{1} \quad
    Pragya Sharma\textsuperscript{2} \quad
    Walid A. Hanafy\textsuperscript{1} \quad
    David Irwin\textsuperscript{1} \quad
    Mani Srivastava\textsuperscript{2} \quad
    Prashant Shenoy\textsuperscript{1} \\
    \textsuperscript{1}University of Massachusetts Amherst \\
    \textsuperscript{2}University of California Los Angeles
}
\renewcommand{\shortauthors}{Shastri et al.}

\begin{abstract}
Foundation models (FMs) are increasingly used as backbones for downstream tasks across language, vision, time-series, and multimodal applications. Yet existing model-serving systems deploy each customized task as an independent model instance, thereby replicating heavyweight backbones, wasting accelerator memory, and losing opportunities to amortize batching and loading costs.
This paper presents \systemName, a serving system that treats FM backbones as a virtualization substrate for deployment sharing. \systemName presents each task with a virtual foundation model (\vFM), a logically private FM instance backed by a shared physical FM. This abstraction lets independently customized tasks share a backbone while preserving task-specific extensions, independent lifecycles, and task-level isolation. In addition, we propose a batch-aware fair-queueing scheduler that combines weighted task-level sharing with inter- and intra-task batching across co-located tasks.
We implement a \systemName-based serving stack spanning task construction, sharing-aware deployment, and runtime execution. Across 7 FM backbones (16 variants) and 92 downstream tasks, \systemName reduces latency by up to 80\% over spatial partitioning and 33.3\% over best-effort co-location, while hosting up to 6$\times$ more tasks at cluster scale.

\end{abstract}

\maketitle

\section{Introduction}\label{sec:intro}
Recent advances in artificial intelligence (AI) have changed how we design and build applications across video analytics, web services, chatbots, AR/VR, recommendation systems, and IoT~\cite{Shen2019:Nexus, Satya2021:TheRole, Siam2025:AIoT}.
Increasingly, these advances are driven by foundation models (FMs), a new generation of general pretrained \emph{backbone models} trained on broad, large-scale datasets using self-supervision and capable of supporting a wide range of \emph{downstream} tasks with limited or no fine-tuning~\cite{Bommasani2021:FoundationModels, Baris2025:FMs-CPS-IoT}.
Foundation models now span many domains and modalities, including time series~\cite{Goswami2024:MOMENT, Ansari2024:Chronos}, vision~\cite{Oquab2024:DinoV2, Radford2021:CLIP}, and natural language~\cite{touvron2023:LLaMA, GEMMA2024, Brown2020:FewShot, Guo2025:DeepSeek}.
Large language models (LLMs) are examples of text-based foundation models, while vision-language models (VLMs)~\cite{Liu2023:LLaVA, Meta2024:Llama32-Vision, Yao2024:MiniCPMV, Wang2024:Qwen2VL} are examples of multimodal foundation models supporting text and vision.

A key benefit of foundation models is that they reduce the need to train separate models for different downstream tasks, while still being \emph{customizable} and \emph{extensible} to specific task needs. For instance, a task using a foundation model will typically use a task-specific head (\eg a classifier head) and can further fine-tune the model using parameter-efficient fine-tuning approaches~\cite{peft}.
Despite the multi-task nature of foundation models, conventional model-serving systems, such as NVIDIA Triton~\cite{nvidia_triton}, are still built around task-specific model deployments. These systems improve resource efficiency through request batching and GPU sharing~\cite{Romero2021:INFaaS, Shen2019:Nexus, Zhang202:MArk, Ahmad2024:Proteus, Daniel2017:Clipper}, but a common theme across prior approaches is the use of \emph{a separate model instance for each task}, which is a natural design choice for task-specific models.

\begin{figure}[t]
    \centering
    \includegraphics[width=\linewidth]{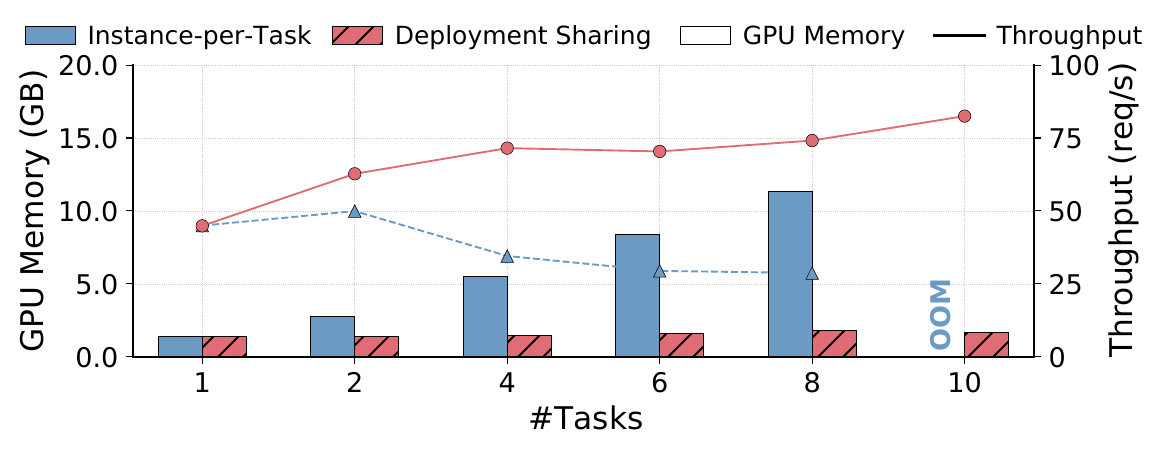}
    \caption{Benefits of FM sharing in terms of memory demand and throughput across a number of tasks and modalities.}
    \Description{Benefits of FM sharing in terms of memory demand and throughput across a number of tasks and modalities.}
    \label{fig:sharing_motivation}
\end{figure}

However, the instance-per-task approach is a poor fit for FMs. FMs are backbone-heavy models designed for multi-task use and are often orders of magnitude larger than task-specific models. As a result, tasks built on the same FM are forced to load redundant copies of the same backbone, incurring duplicate memory usage and poor accelerator utilization. This pattern precludes what we call \emph{deployment sharing}, in which multiple tasks share a common FM instance.
\autoref{fig:sharing_motivation} illustrates the benefits of FM sharing as the number of tasks increases, with requests issued in a closed-loop manner using the settings detailed in \autoref{sec:eval_setup}.
As shown in \autoref{fig:sharing_motivation}, co-locating 10 time-series tasks on a shared MOMENT-large backbone requires only 1.17$\times$ the memory of a single task, compared to 10$\times$ under independent deployment. This savings is possible because the backbone dominates pipeline memory, while task-specific components (decoder heads, adapters, encoders) add only marginal overhead per task. In addition, the figure shows that sharing FMs can increase throughput by 83\%, whereas replicating FMs can decrease it, as different tasks compete for GPU resources.

However, sharing FMs across tasks introduces three key challenges.
\emph{\textbf{Challenge 1:} Deployment sharing introduces cross-task interference.} When multiple tasks share the same FM, their inference requests execute on the same model instance and may be batched together, increasing the risk of cross-task interference.
As a result, load spikes from one task can consume shared resources and increase queueing delays for other tasks.
Although modern GPUs provide device-level virtualization and spatial partitioning mechanisms (\eg NVIDIA MIG~\cite{nvidia_mig}, transparent compute partitioning~\cite{bakita2023:TPCs}, and CUDA Green Contexts~\cite{cuda_green_contexts}), these mechanisms do not directly provide per-task isolation inside a shared FM, where interference arises at the model-serving and request-scheduling layers.
\emph{\textbf{Challenge 2:} Deployment sharing complicates task-specific customization.}
Sharing an FM should not preclude tasks from specializing the FM for task-specific semantics. Tasks may require task-specific heads or fine-tuned parameters, but deployment sharing must separate a task's logical view of a customized model from the physical state of the shared FM. Otherwise, customization by one task could affect all other tasks that use the same FM.
\emph{\textbf{Challenge 3:} Deployment sharing must accommodate independent task lifecycles over a shared FM.}
In deployment sharing, the set of tasks bound to a shared FM evolves at runtime. Tasks may arrive, change their service objectives, or be removed without redeploying the underlying FM. Multi-task learning approaches~\cite{Liu2019:MultiTask, Standley2020:MultiTask} address task plurality at training time but couple the deployed artifact to a predetermined task set, which is incompatible with runtime task churn.
As a result of these challenges, we argue that the backbone of a foundation model should be treated as a shared system substrate, analogous to how operating systems virtualize CPUs or memory, rather than as a per-task deployment artifact.
These challenges call for treating the FM backbone as shared infrastructure rather than a private per-task artifact, mirroring classical virtualization systems that multiplex physical resources across isolated execution contexts~\cite{Popek1974:VMRequirements,Creasy1981:VM370,Merkel2014:Docker,Smith2005:VMs,lxc}.

\begin{figure}
    \centering
    \includegraphics[width=\linewidth]{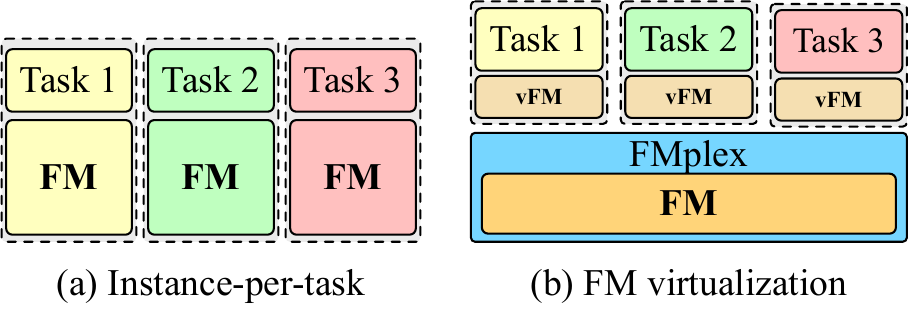}
    \caption{Comparing (a) the instance-per-task approach, where each task loads its own FM and the backbone is replicated, with (b) \textbf{our} FM virtualization approach, where each task is presented with a virtual FM (\vFM) backed by a shared physical FM, enabling deployment sharing.}
    \Description{Comparing (a) the instance-per-task approach, where each task loads its own FM and the backbone is replicated, with (b) \textbf{our} FM virtualization approach, where each task is presented with a virtual FM (\vFM) backed by a shared physical FM, enabling deployment sharing.}
    \label{fig:FM_virtualization_concept}
\end{figure}

In this paper, we present \systemName, a new approach for FM serving that addresses these challenges.
At its core is foundation model virtualization, where \systemName presents each task with a virtual foundation model (\vFM), a logically private FM instance that the task can fully customize or fine-tune, while \systemName multiplexes many \vFMs over shared physical FMs.
By mapping \vFMs to shared physical FMs, \systemName improves resource efficiency through deployment sharing without duplicating the FM. This \vFM abstraction directly addresses Challenges~2 and~3, isolating each task's customization state from the shared physical FM and allowing tasks to be created or removed without redeploying it.
\autoref{fig:FM_virtualization_concept} illustrates the difference between (a) the instance-per-task approach, where tasks are coupled to separate FM instances and do not share a backbone, and (b) our proposed FM virtualization approach, where tasks share the same backbone through \vFMs.
In addition, \systemName isolates \vFMs from one another using Batch-aware Fair Queueing (\texttt{BFQ}), a fair-sharing approach designed for shared FM execution.
Like traditional fair-share schedulers, \texttt{BFQ} assigns each \vFM a weight and allocates shared FM execution in proportion to it, thereby preventing interference among tasks sharing the same physical FM.
However, \texttt{BFQ} is designed for deployment-sharing scenarios, maximizing efficiency through inter- and intra-task request batching.
In doing so, \texttt{BFQ} addresses Challenge~1, providing per-task isolation without sacrificing the efficiency benefits of consolidation.
Finally, we integrate \systemName with \FMAPI (an API for customizing FMs) and \systemEngine (a cluster-level controller for FM sharing-aware deployment and runtime adaptation) to build an end-to-end serving stack for extensible FMs.

In designing and implementing \systemName, we make the following contributions:

\begin{itemize}[leftmargin=*]

\item \textbf{FM Virtualization for Deployment Sharing.}
We introduce \systemName, an FM virtualization layer that presents each task with a logically private \vFM while multiplexing many \vFMs over shared physical FMs. This abstraction decouples task customization and lifecycle management from physical FM deployment; \texttt{BFQ} provides the runtime isolation needed to share a physical FM while preserving the throughput benefits of inter- and intra-task batching.

\item \textbf{Sharing-Based Serving Stack.}
We build an end-to-end serving stack around \systemName, integrating \FMAPI for extensible task construction and \systemEngine for cluster-level sharing-aware deployment. This design makes FM virtualization operational across heterogeneous FMs and multi-task workloads.

\item \textbf{Prototype Implementation.} We implement the \systemName-based serving stack with support for multiple FM modalities (\eg time-series, vision FMs, LLMs, and VLMs) and runtimes (\eg PyTorch and vLLM), demonstrating the flexibility and utility of our design.

\item \textbf{Evaluation.}
We evaluate \systemName across 7 backbone foundation models (16 variants) and 92 downstream tasks spanning time-series, vision, language, and vision-language modalities. For two tasks sharing the same backbone, \systemName reduces latency by up to 33.3\% over current co-location approaches and by 80\% over spatial-partitioning approaches. \texttt{BFQ} sustains 0.97--0.98 fairness at 84\,RPS under asymmetric service weights, where classical fair-share schedulers achieve similar fairness at only 37\,RPS. At cluster scale, \systemName hosts up to 6$\times$ more tasks than current co-location approaches at low load, where memory is the binding constraint, and 8--12\% more at moderate and high load, where compute is the binding constraint.

\end{itemize}

\section{Background}\label{sec:background}
This section provides background on foundation models, FM extension mechanisms, and model-serving systems.

\subsection{Foundation Models}
\label{sec:bg_fm}

Recent advances in AI have shifted model development from task-specific architectures toward general-purpose foundation models. Earlier approaches typically trained specialized models for tasks such as classification, object detection, and segmentation~\cite{resnet, efficient-net, Redmon2016:YOLO, Yang2016:IQA, LSTM}. In contrast, foundation models (FMs) are general-purpose pretrained \emph{backbone models} trained on large, heterogeneous datasets using self-supervision and can be adapted to a wide range of \emph{downstream} tasks~\cite{Bommasani2021:FoundationModels, Baris2025:FMs-CPS-IoT}, achieving strong performance with limited or no task-specific adaptation.
FMs cover multiple input modalities, including time series, vision, language, and combinations of these. Examples include time-series foundation models (TSFMs)~\cite{Goswami2024:MOMENT, Ansari2024:Chronos, feofanov2025:Mantis}, vision foundation models (VFMs)~\cite{Oquab2024:DinoV2, Radford2021:CLIP, Dosovitskiy2021:ViT}, LLMs~\cite{touvron2023:LLaMA, GEMMA2024, Brown2020:FewShot, Guo2025:DeepSeek}, and VLMs~\cite{Liu2023:LLaVA, Meta2024:Llama32-Vision, Yao2024:MiniCPMV, Wang2024:Qwen2VL}. Other FMs target specific application domains, including energy~\cite{Maji2025:CarbonX, Kumar2025:MixForecast}, finance~\cite{shi2025:kronos, zhu2025:FinCast, Rao2025:LLMFinance}, and health~\cite{Gnassounou2025:TSEEG, Huang2025:MedTS, pillai2025:PaPaGei, Dong2024:SEGMedical}.
Across these settings, a single FM often supports several downstream tasks, an ability we refer to as \emph{task sharing}, meaning that the model can be adapted to different tasks even if each task is deployed as a separate model instance. For example, MOMENT~\cite{Goswami2024:MOMENT}, a time-series FM, can perform classification, forecasting, and anomaly detection from a single backbone.

\subsection{Extending Foundation Models}
\label{sec:bg_extending}

\begin{figure}
    \centering
    \includegraphics[width=\linewidth]{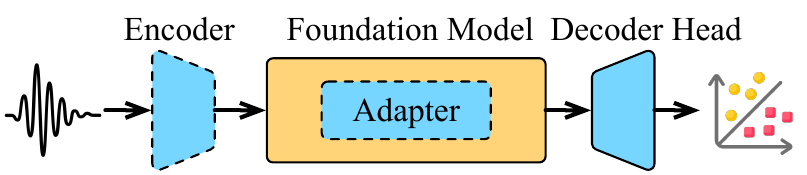}
    \caption{Architecture of an FM-based task pipeline featuring a \emph{decoder head}, optional \emph{encoder} and fine-tuning \emph{adapter}.}
    \Description{The architecture of an FM-based task pipeline featuring a \emph{decoder head}, optional \emph{encoder} and fine-tuning \emph{adapter}.}
    \label{fig:fm_ead}
    \vspace{-0.5em}
\end{figure}

FMs differ in how their backbones process requests. For instance, \emph{representation-based} FMs use the backbone as a feature extractor, with a fixed input and output shape. Examples include time-series FMs such as MOMENT~\cite{Goswami2024:MOMENT} and vision FMs such as DINOv2~\cite{Oquab2024:DinoV2}. In contrast, \emph{token-based} FMs operate autoregressively over token sequences, taking tokenized input and producing tokenized output; examples include LLMs such as LLaMA~\cite{touvron2023:LLaMA} and VLMs such as LLaMA-3.2-Vision~\cite{Meta2024:Llama32-Vision}.
In either case, each task relies on the FM backbone for the shared computation that supports multiple downstream tasks.
The backbone is the primary component, often multiple orders of magnitude larger in parameters and memory than the remaining task-specific components, as shown in \autoref{fig:sharing_motivation}, and is heavyweight to instantiate (large weights, non-trivial loading warmup time). Tasks, by contrast, consist only of a task specification and lightweight extension weights.
\autoref{fig:fm_ead} illustrates the common components of an FM-based task pipeline and the extension points used to customize an FM\footnote{Throughout the paper, we use \emph{encoder} to refer to an input-side adaptation module that pre-processes data before invoking the FM, \emph{decoder head} or \emph{decoder} to refer to a task-specific output head, \emph{adapter} to refer to a lightweight PEFT module attached to the FM backbone, and \emph{pipeline} to refer to the user-defined composition of these components around an FM backbone.}.

To extend an FM, a task typically includes a task-specific \emph{decoder head} attached to backbone outputs. Example heads include MLP classifiers or dense decoders for tasks such as classification, forecasting, detection, and segmentation~\cite{Goswami2024:MOMENT, pillai2025:PaPaGei, Baris2025:FMs-CPS-IoT, Maji2025:CarbonX, Simeone2026:TSFM-Energy}. In addition, rather than using the FM backbone as is, tasks may fine-tune the backbone to improve accuracy. Parameter-efficient fine-tuning (PEFT) adapters~\cite{peft, Balne2024:PEFTAnalysis}, such as LoRA~\cite{Hu2022:LoRA}, enable such fine-tuning while updating only a small subset of parameters within the backbone~\cite{Houlsby2019:PETL, Xu2023:PEFTLLM, Shen2025:EdgeLoRA, Sheng2024:S-LoRA}.
A task may also use an encoder to perform input-side adaptation, transforming or normalizing raw inputs into an intermediate representation before feeding them into the backbone.
Finally, some token-based FMs (\eg LLMs) additionally support prompt-based adaptation, where tasks are expressed through input tokens without modifying backbone weights.

Despite this flexibility, current approaches to extending FMs are heterogeneous, and existing implementations are often ad hoc and model-specific, limiting portability across runtimes.

\subsection{Model-Serving Systems}

Model-serving systems are runtime frameworks for efficient inference of machine learning models.
A model-serving system can serve multiple models, support multiple concurrent tasks, and manage execution across one or more GPUs and servers. Examples include NVIDIA Triton~\cite{nvidia_triton}, vLLM~\cite{Kwon2023:vLLM}, and other inference-serving frameworks~\cite{Romero2021:INFaaS, Daniel2017:Clipper, Ahmad2024:Proteus, Zhang202:MArk, Shen2019:Nexus, ng2024TailClipper2, Gujarati2020:Clockwork, Liang2020:Queuing}.
As discussed in \autoref{sec:intro}, conventional model-serving systems load a separate model instance for each task and use that instance to serve incoming requests.
Foundation models create new opportunities for more resource-efficient execution beyond this instance-per-task approach.
For example, if two tasks use the same backbone but different decoder heads, the serving system can load a single copy of the backbone and share it between the tasks, rather than deploying separate backbone copies.
In this setup, the serving system executes each task's requests (or batch of requests) against the shared backbone and then applies the task-specific decoder head to produce the final output, an approach we denote as \emph{deployment sharing}.
Such deployment sharing can significantly improve resource efficiency, especially by reducing GPU memory footprint. However, it also introduces cross-task interference, task-specific customization, and independent task-lifecycle challenges, which \autoref{sec:requirements} formalizes as design requirements.

\section{Design Overview}\label{sec:overview}
This section presents the four design requirements for sharing-aware model serving, followed by an overview of \systemName and the \systemName-based model serving stack.

\subsection{Design Requirements}\label{sec:requirements}

\systemName must satisfy four design requirements.

\noindent\reqcircle{R1} \textbf{Backbone sharing as a first-class primitive.} FM backbones are heavyweight system objects, often requiring gigabytes of GPU memory and substantial loading and warmup time, whereas task-specific state (e.g., decoder weights, PEFT adapters, queue state, and accounting metadata) is comparatively lightweight.
The serving system must support multiple tasks per FM instance, treating the backbone as a shared substrate rather than a per-task resource.

\noindent\reqcircle{R2} \textbf{Decoupling backbone from task lifecycle.}
The serving system must therefore decouple the backbone-instance lifecycle from the task lifecycle, allowing tasks to maintain independent queues, customization state, and accounting identities independent of the specific backbone instance currently serving them. This decoupling enables runtime adaptation through lightweight task-state movement and rebinding across existing FM instances, rather than requiring expensive FM reloading or full model redeployment.

\noindent\reqcircle{R3} \textbf{Task-level isolation.} Tasks have heterogeneous request rates and arrival patterns.
The serving system must provide a per-task isolation boundary, a named, addressable context with its own request queue and accounting state, at which scheduling, fairness, and admission decisions are enforced.

\noindent\reqcircle{R4} \textbf{Unified extensibility.} FMs span a wide range of architectures, modalities, and processing mechanisms.
The serving system must expose a uniform extensibility interface through which tasks attach decoders, adapters, and prompts to a backbone, without per-task changes to the runtime or per-backbone reimplementation of the extension layer.

\subsection{\systemName Overview}\label{sec:design_overview}

\begin{figure}
    \centering
    \includegraphics[width=0.8\linewidth]{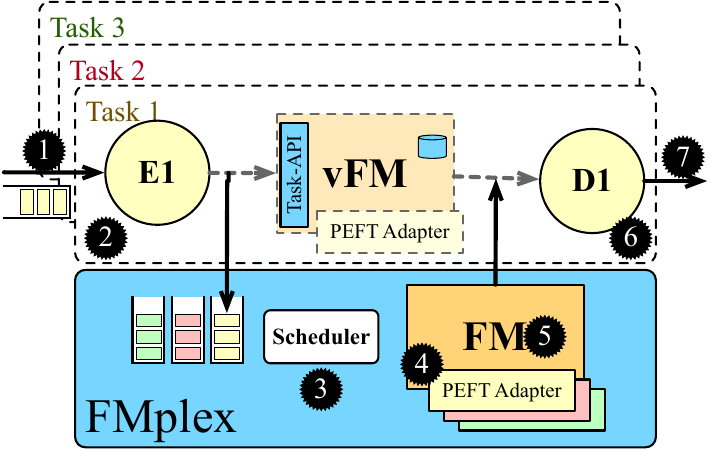}
    \caption{Overview of \systemName. The figure contrasts the task-side \vFM view with the underlying physical FM deployment. Steps 1--7 illustrate the request path through \systemName.}
    \Description{Overview of \systemName. The figure contrasts the task-side \vFM view with the underlying physical FM deployment. Steps 1--7 illustrate the request path through \systemName.}
    \label{fig:FMplex_arch}
\end{figure}

\autoref{fig:FMplex_arch} depicts the overview of \systemName. At a high level, \systemName decouples each task's logical view of the foundation model from its physical substrate. Analogous to a Hypervisor~\cite{Smith2005:VMs} or Containers~\cite{Merkel2014:Docker, lxc}, \systemName presents each task with a virtual foundation model (\vFM) and multiplexes many \vFMs over a single shared physical FM.
\systemName comprises three components that jointly realize \reqcircle{R1}--\reqcircle{R4}. The \textit{\vFM abstraction} (\autoref{sec:fm_virt_vfm}) realizes this per-task decoupling, satisfying \reqcircle{R1} and \reqcircle{R2}, giving the task its own invocation context, state, and service policy, and letting the task extend the \vFM using FM customization approaches.
The \textit{batch-aware fair-share scheduler} (\autoref{sec:design_isolation}) arbitrates execution across colocated \vFMs sharing a backbone, satisfying \reqcircle{R3}.
\textit{\FMAPI} (\autoref{sec:design_FMAPI}) is the task-facing programming interface for composing encoders, decoders, and adapters around a \vFM, satisfying \reqcircle{R4}.
To the best of our knowledge, \systemName is the first system to treat foundation-model backbones as a virtualization substrate, extending these classical primitives to FM serving.

\noindent\textbf{Sharing-Based Serving Stack.}
\autoref{fig:system_overview} presents the end-to-end serving stack built on top of \systemName. Users construct task pipelines through \FMAPI's client-side library and submit them as deployment artifacts to \systemEngine. \systemEngine is a sharing-aware deployment manager that maintains the current deployment state, mapping tasks to physical FM instances across the cluster based on per-task profiles, SLOs, and observed load. We detail \systemEngine's sharing-aware deployment and elastic-adaptation mechanisms in \autoref{sec:orchestration}.

\section{FM Virtualization}\label{sec:design}
This section describes the design of \systemName, our virtualization approach, and its three components: the \vFM abstraction (\autoref{sec:fm_virt_vfm}), the batch-aware fair-share scheduler (\autoref{sec:design_isolation}), and the \FMAPI programming interface (\autoref{sec:design_FMAPI}).

\subsection{Virtual Foundation Model (\vFM)}\label{sec:fm_virt_vfm}

The key abstraction exported by \systemName is the \emph{virtual foundation model} (\vFM), which presents each task with the illusion that it is running exclusively on its own foundation model.
The \vFM decouples each task from the underlying physical FM and provides a control point for accounting, isolation, and policy enforcement.
Without this abstraction, tasks would appear to the runtime only as anonymous request streams entering a shared backbone, making task-specific customization, per-task isolation, and accounting impossible to enforce. A \vFM includes three facets:

\begin{itemize}[leftmargin=*]

    \item \emph{Virtual Queue.} A task invokes its \vFM as if it were a dedicated foundation model. \systemName intercepts each invocation and routes it through a per-task virtual queue, redirecting execution onto the shared backbone without the task interacting with it directly.

    \item \emph{Task Extensions.} Each \vFM holds the task-specific extensions (encoder, decoder, and PEFT adapter) that customize the shared backbone for the task's downstream behavior.

    \item \emph{State and accounting.} Each \vFM holds task-level configuration (SLOs, priorities, fair-share weight) and has its own named accounting identity. \systemName tracks resource usage, request rates, and admitted load per \vFM, and uses this information to drive admission control, fair sharing, and SLO enforcement.

\end{itemize}

\autoref{fig:FMplex_arch} traces an end-to-end request through \systemName. A request first arrives at the task endpoint (\eg via HTTP)~\circled{1} and enters the task-specific frontend pipeline~\circled{2}. When the task invokes the \vFM via the \FMAPI~\circled{3}, \systemName redirects the call to the task's virtual queue~\circled{4}. The scheduler then selects queued requests for execution on the shared FM~\circled{5} and binds task-specific adapters as needed~\circled{6}. The resulting outputs are returned to the task pipeline, which produces the final response~\circled{7}.

\subsection{Batch-aware Fair Sharing and Isolation}\label{sec:design_isolation}

The \systemName scheduler realizes fair sharing and isolation across \vFMs, providing each task a predictable share of execution capacity, independent of the behavior of co-resident tasks. However, this isolation should not erode the efficiency gains of inter- and intra-task batching. Meeting both requirements is challenging because FM scheduling differs from conventional CPU or GPU scheduling in four ways.
First, fair-share scheduling approaches~\cite{STFQ,WFQ, waldspurger1994lottery,GPS}, focus on per-request basis, whereas model-serving systems rely heavily on request batching to improve throughput.
Second, batching improves accelerator utilization but increases queueing delay due to head-of-line blocking~\cite{Daniel2017:Clipper, Gujarati2020:Clockwork, Romero2021:INFaaS}. Third, PEFT adapters introduce batching compatibility constraints and switching overheads that existing fair schedulers do not capture~\cite{STFQ,WFQ,Liang2023:ETF}. Fourth, \systemName must support both representation-based FMs (\eg vision FMs with request-level batching) and token-based FMs (\eg LLMs with continuous batching and long-lived decode iterations)~\cite{Yu2022:Orca, Kwon2023:vLLM}.
To address these challenges, \systemName introduces \texttt{BFQ}, a batch-aware fair queueing scheduler. \texttt{BFQ} determines which \vFMs to serve, how many requests to admit from each \vFM, which requests can be co-batched, and how execution is sequenced over time.

\noindent\textbf{Batch-aware Fair Scheduling.}
\texttt{BFQ} is a work-conserving scheduler for shared FM execution that approximates weighted fair sharing while accounting for batching efficiency. We describe its representation-based implementation first, then explain how the same service-time accounting applies to token-based runtimes.
\texttt{BFQ} is derived from start-time fair queueing (\texttt{STFQ})~\cite{STFQ}, but extends it from per-request ordering to batch formation. For task $i$ with weight $w_i$, request $j$ is assigned a start tag $S_i^j$ and a finish tag $F_i^j$:
{\footnotesize
\begin{align}
    S_i^j &= \max\{F_i^{j-1}, v\}, \\
    F_i^j &= S_i^j + \frac{l}{w_i},
\end{align}}
where $v$ is the scheduler's global tag at arrival, computed as $v = \max_i F_i^{\text{last}}$, the maximum finish tag across each task's most recently dispatched request, and $l$ is the expected service time of a single request. Unlike \texttt{STFQ}, \texttt{BFQ} must jointly decide \emph{which} requests to serve and \emph{how many} to batch.
It therefore orders queued requests by start tag and iteratively extends the candidate batch until either (1) the batch reaches maximum allowed batch size $B_{\max}$, typically configured at the FM's profiled throughput knee, beyond which larger batches yield limited throughput gains (see \autoref{fig:sharing_motivation}), or (2) admitting another request would push any selected request past its SLO deadline given the batch's estimated completion time.
Finally, to account for adapter-compatibility constraints, \texttt{BFQ} batches the requests across the common backbone first then integrate adapter difference sequentially across in-compatible requests thereby respecting adapter compatibility while preserving \texttt{BFQ}'s fair-share accounting.

\begin{figure}[t]
    \centering
    \includegraphics[width=\linewidth]{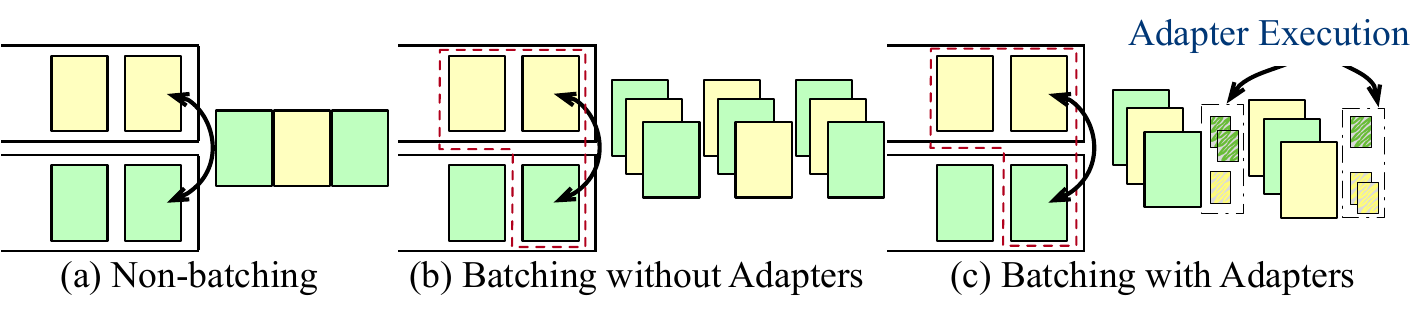}
    \caption{\texttt{BFQ} behavior under different scenarios.}
    \Description{\texttt{BFQ} behavior under different scenarios.}
    \label{fig:fairness_example}
\end{figure}

\begin{table}[t]
    \centering
    \caption{High-level Description of \FMAPI}
    \label{tab:FMAPI}
    \resizebox{\linewidth}{!}{
    \begin{tabular}{r|c|l}
    \toprule
        \textbf{Call} & \textbf{Input} & \textbf{Description} \\ \midrule
        \texttt{\textcolor{blue}{class} Encoder:} & None & Encoder Interface \\
        \texttt{E.run(..)} &request & Process a request in Encoder\\\midrule
        \texttt{\textcolor{blue}{class} Decoder:} &None& Decoder Interface \\
        \texttt{D.run(..)} &request& Process a request in Decoder\\\midrule

        \texttt{\textcolor{blue}{class} Adapter:} &None& Adapter Interface \\\midrule

        \texttt{\textcolor{blue}{class} \vFM:} & FM-ID & \vFM Interface \\
        \texttt{v.run(..)} &request& Process a request in an FM\\\midrule

        \texttt{\textcolor{blue}{class} Pipeline:} &\vFM& A task pipeline with FM as input\\
         \texttt{p.add\_encoder()} &Encoder& Add encoder to pipeline \\
         \texttt{p.add\_decoder(..)} &Decoder& Add decoder to pipeline \\
         \texttt{p.attach\_adapter(..)} & PEFT Adapter& Attach an adapter to \vFM \\
         \texttt{p.remove\_adapter(..)} & Adapter-ID & Remove an adapter from \vFM \\
         \texttt{p.run(..)} &request& Process a request in pipeline\\\midrule
         \texttt{p.train(..)} & Data Loader, Hyper Parameters & Train encoder, adapter, or decoder
         \\\bottomrule
    \end{tabular}
    }
\end{table}

Batching introduces a \emph{mismatch} between the expected and realized service times. The reason is that when requests are co-batched, the amortized service time per request decreases, so the start and finish tags computed under single-request service overestimate the actual service consumed. To correct this mismatch, after executing a batch, \texttt{BFQ} recomputes the finish and start tags of every queued request whose task contributed to the executed batch, using a batch-dependent service time:
{\footnotesize
\begin{equation}
    F_i^j = S_i^j + \frac{l_i(b)}{w_i},
\end{equation}}
where $l_i(b)$ denotes the effective per-request service time for task $i$ when one of its requests is co-executed in a compatible sub-batch of size $b$.
This preserves batching efficiency while keeping the schedule informed with realized service times.

\noindent\textbf{Token-based FMs.}
\systemName applies the same accounting principle to token-based FMs by charging token-level work in service-time units. This keeps \vFM scheduling consistent across request-level and token-level runtimes without introducing a new token-scheduling policy.

\noindent\textbf{Example.}
\autoref{fig:fairness_example} illustrates three cases. In (a) requests are not batched \eg if batching would violate the SLO. In (b), batching preserves the SLO, so the scheduler co-batches requests.  In (c),
adapter incompatibility prevents batching across adapter fused backbones. Hence we batch across single base backbone  and loop across delta caused due to adapters.

\subsection{\FMAPI}\label{sec:design_FMAPI}

\FMAPI is the task-facing realization of the \vFM abstraction (\autoref{sec:fm_virt_vfm}).
It gives task developers a small interface for specifying task-local extensions while giving \systemName the metadata needed to instantiate, bind, and manage those tasks over shared FM backbones.
\autoref{tab:FMAPI} summarizes the high-level API abstractions across encoders, decoders, and \vFMs. A task pipeline is constructed around a \vFM, to which encoders, decoders, and PEFT adapters (\eg LoRA) are attached.
The API provides primitives such as \texttt{add\_encoder()}, \texttt{add\_decoder()}, or attaching adapters to the \vFM via \texttt{attach\_adapter()}, which is applied internally during the forward pass.
\autoref{lst:moment_example} illustrates a task pipeline based on MOMENT-Base~\cite{Goswami2024:MOMENT}, composed with a \texttt{LinearChannelCombiner} encoder, an \texttt{MLP} decoder, and a \texttt{LoRA} adapter.

\noindent\textbf{Fine-tuning.}
Beyond pipeline specification, \FMAPI supports end-to-end fine-tuning of task pipelines, where the \vFM backbone is treated as frozen by default, while encoder, decoder, and adapter weights are trainable via the same pipeline object used for serving. Using the same pipeline object for fine-tuning and serving lets \systemName package the resulting task state as a deployment artifact without changing the serving runtime. \autoref{lst:moment_full} in the appendix shows a complete training pipeline using \FMAPI.

\noindent\textbf{Deployment Artifacts.}
A completed task pipeline is packaged as a deployment artifact consumable by \systemName. This artifact includes the pipeline specification, encoder, adapter, and decoder implementations, their associated weights, and metadata describing the task's runtime and load requirements. These artifacts allow \systemName to instantiate the task, bind it to the appropriate FM, and manage it within the FM serving infrastructure.

\lstset{basicstyle=\small\ttfamily,columns=fullflexible}
\begin{listing}[t]
\centering
\begin{minted}[fontsize=\small, linenos, numbersep=8pt, xleftmargin=1.6em, breaklines=true]{python}
P=Pipeline(vMomentModelBase(...))
P.add_encoder(LinearChannelCombiner(...))
P.add_decoder(MLP(...))
P.attach_adapter(LoRA_Adapter(...))
res=P.run(...)
\end{minted}
\centering\caption{High-level example of a task pipeline attaching Moment-Base FM with MLP decoder, Linear channel encoder, and LoRA adapter.}
\label{lst:moment_example}
\end{listing}

\section{\systemName-based Serving Stack}\label{sec:orchestration}
\begin{figure}[t]
    \centering
    \includegraphics[width=\linewidth]{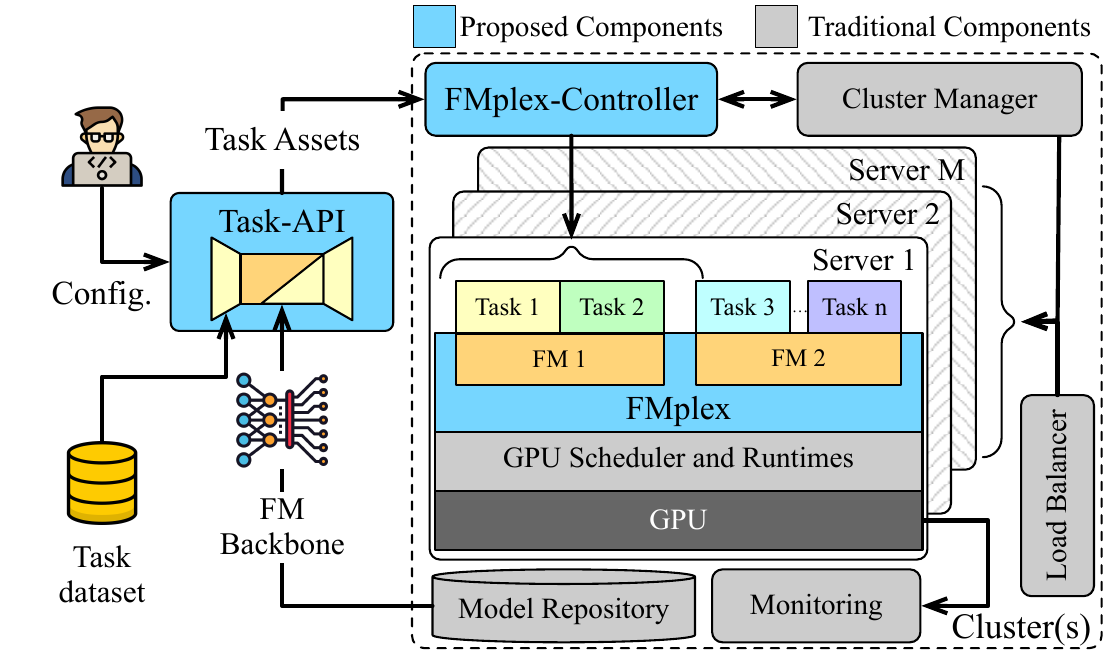}
    \caption{End-to-end serving stack on top of \systemName.}
    \Description{End-to-end serving stack on top of \systemName.}
    \label{fig:system_overview}
\end{figure}

In this section, we describe the \systemName-based serving stack, which integrates \FMAPI, per-server \systemName instances, and \systemEngine to support task deployment, routing, and adaptation across a cluster.

\subsection{Overview}\label{sec:orch_overview}
The mechanisms in \autoref{sec:design} define how a single server virtualizes shared FM execution through \vFMs, task-local queues, and \texttt{BFQ}. \autoref{fig:system_overview} shows how \systemName fits into a broader model-serving stack. The stack retains standard components such as task endpoints, load balancing, monitoring, and per-server runtimes, but changes the deployment unit from an opaque model instance to a task pipeline bound to a physical FM instance.
\FMAPI packages task pipelines as deployment artifacts, and per-server \systemName instances bind their \vFMs to physical FM instances and execute requests through \texttt{BFQ}. \systemEngine maintains the global deployment plan, deciding which FM backbones to instantiate, which tasks to bind to each backbone, and how routing should adapt as demand changes. This separation between task placement and backbone placement motivates the sharing-aware deployment policy in \autoref{sec:orch_placement} and the elastic adaptation mechanisms in \autoref{sec:orch_elastic}.

\subsection{Sharing-aware Deployment}\label{sec:orch_placement}

\systemEngine supports the virtualization layer by deciding how tasks are distributed among available backbones and, when needed, where new backbones should be instantiated.
To demonstrate the benefits of deployment sharing for model-serving orchestration, we intentionally use a lightweight greedy heuristic rather than a full placement optimizer (\autoref{alg:orchestrate}).
The heuristic prioritizes reusing existing FM backbones and accepts a placement when the profiled memory and compute constraints indicate that the new and existing tasks can meet their SLOs.
Given a task's required backbone, expected demand, SLO constraints, and available server capacity, \autoref{alg:orchestrate} computes a deployment plan that specifies task-to-backbone bindings, task replication across servers, and request routing across active deployments.
This choice exposes the deployment decisions enabled by FM virtualization and isolates the benefits of virtualization from orthogonal cluster-placement optimizations.

\RestyleAlgo{ruled}
\begin{algorithm}[t]
    \footnotesize
    \caption{\texttt{Max-Share Heuristic}}
    \label{alg:orchestrate}
    \KwIn{Task $\mathcal{T}$, servers $\mathcal{S}$.}
    \KwOut{Deployment plan $P$ or $\bot$.}

    $C \gets \emptyset$\tcp*{Candidate servers}
    $D \gets$\texttt{active\_deployments(}$\mathcal{S}, \mathcal{T}$\texttt{)}\;

    \For{$d \in$ \texttt{best\_fit\_order(}$D, \mathcal{T}$\texttt{)}}{
        $C \gets C \cup \{d\}$\tcp*{Prefer existing backbones}
        $P \gets$\texttt{plan(}$\mathcal{T}, C$\texttt{)}\;
        \If{\texttt{feasible(}$P$\texttt{)}}{
            \Return \texttt{commit(}$P$\texttt{)}\;
        }
    }

    \For{$s \in$ \texttt{best\_fit\_servers(}$\mathcal{S}, \mathcal{T}$\texttt{)}}{
        $d \gets$ \texttt{deploy(}$\mathcal{T}.\texttt{FM}, s$\texttt{)}\tcp*{Provision only as needed}
        $C \gets C \cup \{d\}$\;
        $P \gets$\texttt{plan(}$\mathcal{T}, C$\texttt{)}\;
        \If{\texttt{feasible(}$P$\texttt{)}}{
            \Return \texttt{commit(}$P$\texttt{)}\;
        }
    }

    \Return $\bot$\;
\end{algorithm}

\autoref{alg:orchestrate} shows the Max-Share heuristic. \systemEngine first considers active deployments whose backbones can serve the task. If multiple deployments or backbone choices are available, \texttt{best\_fit\_order()} ranks them by how well they can absorb the task while leaving minimal unused capacity. \texttt{plan()} then computes the task-to-backbone bindings and routing assignments over the selected candidate set.
A plan is feasible if any new backbone fits in GPU memory and the selected deployments have enough compute cycles to accommodate the assigned load without violating the SLOs of the new task or existing tasks already sharing those backbones. If no feasible plan exists using active deployments, the heuristic considers servers that can host the required backbone in best-fit order and stops at the first feasible plan.

\subsection{Elastic Adaptation}\label{sec:orch_elastic}

FM virtualization also gives \systemEngine a cheaper response to some load changes and failures. Existing serving systems commonly trigger scaling from monitored load or failures~\cite{Daniel2017:Clipper, Romero2021:INFaaS, Gujarati2020:Clockwork}, but the response is typically to create or move a complete model deployment. In the \systemName-based serving stack, when a compatible backbone is already running with spare capacity, \systemEngine can instead update the affected task's \vFM binding and routing entry, moving only task-local state such as queue metadata, decoder or adapter state, and scheduler weights. If no compatible backbone is available, the system falls back to normal backbone provisioning. We use this simple rebinding path only to illustrate the operational implication of decoupling tasks from backbones, and evaluate its cost in \autoref{sec:eval_adaptation}.

\section{Implementation}\label{sec:implementation}
Our prototype is implemented in Python in approximately 8K lines of code, integrating three components of the \systemName-based serving stack: \FMAPI, per-server \systemName instances, and \systemEngine. The components communicate through gRPC~\cite{grpc-framework}, which we also use for task submission, deployment control, client requests, load balancing, and routing updates. We will release the code as open source upon acceptance.

\noindent\textbf{\FMAPI.}
\FMAPI is implemented as a Python library for constructing, training, and packaging task pipelines. It builds on PyTorch~\cite{Paszke2019:Pytorch} for encoder, decoder, and backbone execution, and uses PEFT~\cite{peft} for LoRA adapters. The client-side library lets users define and fine-tune pipelines around a \vFM, while the server-side library loads the resulting deployment artifact and exposes the serving-time hooks used by \systemName. Each artifact contains the pipeline specification, task extension code, extension weights, and metadata used by \systemEngine for placement and admission.

\noindent\textbf{Per-server \systemName.}
Each server runs a \systemName instance that maintains the local \vFM registry, task-specific queues, scheduler state, and bindings from \vFMs to physical FM instances. Our prototype uses PyTorch~\cite{Paszke2019:Pytorch} for the main runtime and includes a vLLM integration~\cite{Kwon2023:vLLM} to show how the \vFM boundary can be mapped onto token-based runtimes. Task-specific queues create a per-task scheduling and accounting boundary between task pipelines and shared FMs. We complement this with FM-level performance isolation through spatial GPU partitioning. When multiple FM instances share a single accelerator, \systemName assigns their CUDA streams to disjoint Thread Processing Cluster (TPC) subsets~\cite{bakita2023:TPCs}.

\noindent\textbf{\systemEngine.}
\systemEngine maintains the cluster-level deployment plan and pushes deployment and routing updates to per-server \systemName instances. Updates include task-to-FM bindings, replica deployment, and routing assignments computed by the Max-Share heuristic (\autoref{sec:orch_placement}).

\noindent\textbf{Monitoring and Profiling.}
At task registration, \systemName runs a lightweight profiling pass. FM-level estimates (backbone memory, loading time, and service time as a function of batch size) are computed once per backbone and reused across all tasks bound to it. Only task-specific overhead (encoder, decoder, and adapter loading/unloading) is profiled per task.

\section{Evaluation}\label{sec:evaluation}
We first quantify the resource and performance benefits of backbone sharing, then show that \systemName preserves per-task service shares and isolation under contention. We next study \systemName at cluster scale. We close with microbenchmarks, covering adaptation and extension overhead, and a discussion of the strengths and limitations of our approach. The evaluation addresses the following research questions.

\begin{itemize}[leftmargin=*]
    \item What are the benefits of backbone sharing compared to current deployment strategies? \reqcircle{R1} \reqcircle{R2}
    \item Does \systemName preserve per-task shares and performance isolation under contention? \reqcircle{R3}
    \item Does \FMAPI support diverse FM modalities and extension mechanisms? \reqcircle{R4}
\end{itemize}

\subsection{Experimental Settings}\label{sec:eval_setup}
\subsubsection{Experimental Testbed.}
We deploy \systemName on a cluster of up to 16 AWS \texttt{g4dn.2xlarge} instances, each equipped with a single NVIDIA T4 GPU (16\,GB VRAM, 2560 CUDA cores), 8 vCPUs, and 32\,GB of system memory. All instances share a common NFS namespace via Amazon EFS for model weights and task artifacts.

\subsubsection{Foundation Models.}
We evaluate \systemName across 7 backbone FMs (16 variants in total) and 92 downstream tasks (classification, regression, \etc) spanning time-series, vision, language, and vision-language modalities. \autoref{tab:FM_backbones} summarizes the backbones used and the number of downstream tasks built on each. This diversity exercises \systemName's virtualization abstraction across heterogeneous FM families and extension mechanisms.

\begin{table}[htbp]
\centering
\caption{Foundation models used in our evaluation.}
\label{tab:FM_backbones}
\setlength{\tabcolsep}{4pt}
\footnotesize
\resizebox{\linewidth}{!}{%
\begin{tabular}{llll}
\toprule
\textbf{Modality} & \textbf{Backbone} & \textbf{Sizes} & \textbf{Tasks} \\
\midrule
\multirow{2}{*}{Time Series}
  & MOMENT~\cite{Goswami2024:MOMENT}   & (S/B/L)   & \multirow{2}{*}{%
      \shortstack[l]{classification (2), \\ forecasting (5),
      regression (3)}
    }\\
  & Papagei~\cite{Ansari2024:Chronos}   & (P/S/SVRI) & \\
\midrule
\multirow{2}{*}{Vision}
  & DINOv2~\cite{Oquab2024:DinoV2}      & (S/B/L) & \multirow{2}{*}{%
      \shortstack[l]{classification (2), \\regression (1),
      segmentation (1)}
    }\\
  & Swin~\cite{liu2021:SwinTransformer}                & (T/S/B/L) & \\
\midrule
\multirow{2}{*}{VLM}
  & Qwen2.5~\cite{Wang2024:Qwen2VL}    & 3B & \multirow{2}{*}{%
      \shortstack[l]{classification (1)}
    }\\
  & Mistral-7B~\cite{jiang2023:mistral7b} & 7B & \\
\midrule
\multirow{1}{*}{LLM}
  & Qwen2-VL~\cite{Wang2024:Qwen2VL}    & 2B & OCR (1) \\
\midrule
\multicolumn{2}{r}{Total} & 16 & 92\\
\bottomrule
\end{tabular}
}
\end{table}

\subsubsection{Workload Traces.}
Our evaluation uses the following workload traces.
(i)~\textit{Poisson Distribution.} We generate synthetic traces with Poisson-distributed arrivals, sweeping the per-task request rate from 1 to 20 RPS.
(ii)~\textit{Azure Functions Trace~\cite{Shahrad2020:ServerlessWild}.}
We use the Functions trace, a widely used trace in the model serving literature~\cite{Romero2021:INFaaS, Zhang202:MArk, Shen2019:Nexus}. We treat function demand as task demand and cluster them into low-load (6-60\,RPM), moderate-load (60-600\,RPM), and high-load (600-1800\,RPM) groups, randomly select a number of tasks, and assign them to backbones for every experiment, as we will detail in~\autoref{sec:eval_large_scale}.

\subsubsection{Evaluation Metrics}
Our evaluation focuses on five metrics:
(i) \textit{Latency.} We report per-task, per-FM, and per-cluster latencies.
(ii) \textit{Throughput.} We report per-task and aggregate throughput across tasks.
(iii) \textit{Loading Time.} Time to load tasks and FMs, as an indicator of responsiveness.
(iv) \textit{Memory Usage.} GPU memory consumed across all deployments, capturing the resource efficiency of deployment sharing.
(v) \textit{Fairness~\cite{fairness_metric}.} We report fairness to demonstrate the effectiveness of our isolation approach.
\subsubsection{Baselines.}
We compare \systemName against the following baselines\footnote{Existing adapter-sharing systems~\cite{Shen2025:EdgeLoRA, Sheng2024:S-LoRA, Sheng2024:LLMFair} are specialized to LoRA-based LLM serving and do not expose the generalized virtualization, lifecycle decoupling, or heterogeneous FM support that \systemName targets. We therefore compare against deployment and scheduling baselines representing current deployment practice.}.
\begin{itemize}[leftmargin=*]
    \item \textbf{Single Task (\SingleTask)}. Each task runs in isolation on a dedicated GPU, with no sharing or co-location, thereby representing a lower bound on per-task latency achievable on our hardware.
    \item \textbf{Best-Effort Co-location (\BestEffort)}. Multiple task instances share a GPU without any isolation mechanism. Each task loads its own copy of the backbone and competes for compute resources without spatial or temporal partitioning. As earlier model-serving systems did not consider deployment sharing, this baseline represents the deployment model used by existing serving frameworks~\cite{Romero2021:INFaaS, Shen2019:Nexus}.
    \item \textbf{Spatial Partitioning (\SpatialPartition)}. Another no-sharing baseline that statically partitions the GPU's resources among tasks. We rely on \cite{bakita2023:TPCs} to assign tasks to an exclusive, non-overlapping subset of the GPU's Thread Processing Clusters (TPCs), providing spatial isolation while allowing multiple models to reside on a single GPU simultaneously.
    \item \textbf{Shared Backbone Best Effort (\SharedBE)}. Tasks share a single backbone instance but the runtime batches requests in arrival order without any fair-sharing mechanism.
    \item \textbf{Shared Backbone STFQ (\SharedSTFQ)}. Multiple tasks share a single backbone instance with classical Start-Time Fair Queueing~\cite{STFQ}, but without batch-aware request grouping.
\end{itemize}
\subsection{Benefits of FM Sharing}\label{sec:eval_sharing}

\begin{figure}[htbp]
    \centering
    \begin{subfigure}[t]{0.49\columnwidth}
        \centering
        \includegraphics[width=\linewidth]{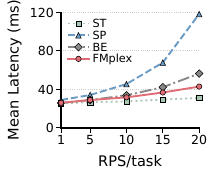}
        \caption{Mean Latency}
        \label{fig:two_tasks_ML_mean}
    \end{subfigure}
    \hfill
    \begin{subfigure}[t]{0.49\columnwidth}
        \centering
        \includegraphics[width=\linewidth]{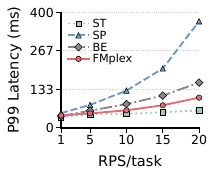}
        \caption{P99 Latency}
        \label{fig:two_tasks_ML_p99}
    \end{subfigure}
    \caption{Comparing \systemName when serving two tasks using Moment-Large ECG and gesture classification tasks across scheduling approaches.}
    \Description{Comparing \systemName when serving two tasks using Moment-Large ECG and gesture classification tasks across scheduling approaches.}
    \label{fig:two_tasks_ML}
\end{figure}

\begin{figure}[htbp]
    \centering
    \begin{subfigure}[t]{0.49\columnwidth}
        \centering
        \includegraphics[width=\linewidth]{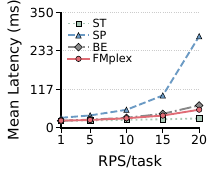}
        \caption{DINOv2-Base}
        \label{fig:two_tasks_models_dino}
    \end{subfigure}
    \hfill
        \begin{subfigure}[t]{0.49\columnwidth}
        \centering
        \includegraphics[width=\linewidth]{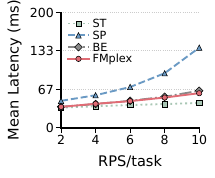}
        \caption{Swin-Large}
        \label{fig:two_tasks_models_swin}
    \end{subfigure}
    \caption{Comparing \systemName when serving two tasks across models.}
    \Description{Comparing \systemName when serving two tasks across models.}
    \label{fig:two_tasks_models}
\end{figure}

This section demonstrates the performance and resource-efficiency benefits of FM backbone sharing, showing how deployment sharing amortizes backbone memory and execution across tasks and how task customizations affect batching behavior.

\subsubsection{Benefits of FM-sharing on Performance}
We first demonstrate the latency benefits of FM sharing relative to the deployment baselines \BestEffort and \SpatialPartition, and quantify the sharing overhead against the per-task latency under no sharing (\ie \SingleTask). \autoref{fig:two_tasks_ML} reports the mean and p99 latency when we co-locate \texttt{ecg\_classification} and \texttt{gesture\_classification} on MOMENT-Large and sweep per-task arrival rate at 1, 5, 10, and 20\,RPS using Poisson inter-arrival times. \autoref{fig:two_tasks_ML_CDF} in \autoref{app:fm_sharing} shows the full latency CDFs across request rates.

As shown, at low load (1\,RPS), \systemName and \BestEffort are comparable to \SingleTask, whereas \SpatialPartition has 13.7\% and 39.4\% higher mean and p99 latency because splitting the GPU inflates per-request service time, even without queueing pressure.
At moderate load (10\,RPS), \SpatialPartition reaches a 45.5 ms mean latency as GPU capacity reduction compounds with queue buildup, and \BestEffort begins to exhibit higher latency. \systemName, however, exhibits latency within 4 ms of \SingleTask and achieving 30.2\% and 6.3\% lower mean (52.8\% and 26.2\% lower p99) latency than \SpatialPartition and \BestEffort, respectively.
At high load (20\,RPS), \SpatialPartition increases to 118.1 ms mean latency (368.7 ms at p99), and \BestEffort to 56.4 ms mean latency (156.9 ms at p99). In contrast, \systemName remains within 12 ms of \SingleTask, widening the mean latency gap to 63.7\% and 23.9\% (71.6\% and 33.3\% w.r.t p99) over \SpatialPartition and \BestEffort, which widens against \BestEffort because it lacks inter-task batching, leaving the backbone's batching capacity unused.

These findings generalize across modalities (\autoref{fig:two_tasks_models}). On DINOv2-Base (\autoref{fig:two_tasks_models_dino}), \systemName reduces mean latency by up to 80\% and 19\% over \SpatialPartition and \BestEffort across a 1--20\,RPS sweep. Similarly, on Swin-Large (\autoref{fig:two_tasks_models_swin}), \systemName achieves up to 57\% and 7\% reductions over the same baselines across a 2--10\,RPS sweep. \autoref{fig:two_tasks_dinobase_CDF} and \autoref{fig:two_tasks_swinlarge_CDF} in \autoref{app:fm_sharing} show the full latency CDFs for the two backbones.
These gains follow from how each approach uses the GPU and schedules requests. \systemName forms cross-task batches over a single backbone, amortizing per-request compute across co-resident tasks. In contrast, \BestEffort runs two backbone replicas that compete for the GPU without coordination, thereby missing many inter-task batched opportunities. In addition, \SpatialPartition spatially partitions the GPU, reducing throughput and overall system efficiency, as we further quantify in \autoref{sec:eval_isolation}.

\begin{figure}[t]
    \centering
    \includegraphics[width=0.5\linewidth]{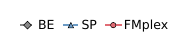}
    \subfloat[\centering 5 RPS]{
    \includegraphics[width=0.49\linewidth]{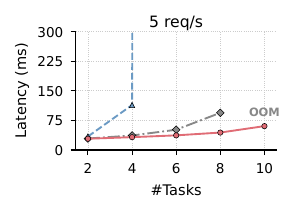}
        \label{fig:single_server_ML_5}
    }
    ~
    \subfloat[\centering 7 RPS]{
        \includegraphics[width=0.49\linewidth]{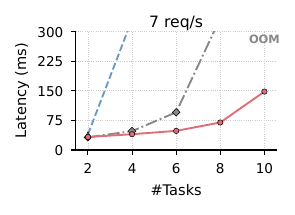}
        \label{fig:single_server_ML_10}
    }
    \caption{Mean latency vs.\ number of co-located tasks on MOMENT-Large, at 5 and 7\,RPS per task.}
    \Description{Mean latency vs.\ number of co-located tasks on MOMENT-Large, at 5 and 7\,RPS per task.}
    \label{fig:single_server_ML}
\end{figure}

\begin{figure}[t]
    \centering
    \includegraphics[width=0.5\linewidth]{figures/Eval/benefits_sharing/increase_num_tasks/ntasks_mean_latency_legend.pdf}
    \subfloat[\centering DINOv2-Base]{
    \includegraphics[width=0.49\linewidth]{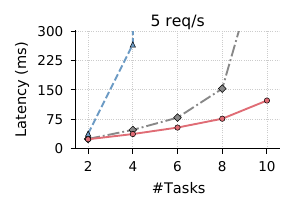}
        \label{fig:single_server_models_ntasks_dino}

    }
    ~
    \subfloat[\centering Swin-Large]{
        \includegraphics[width=0.49\linewidth]{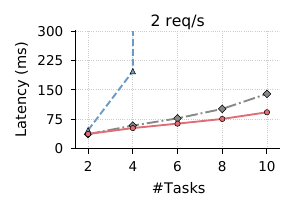}
        \label{fig:single_server_models_ntasks_swin}
    }
    \caption{Mean latency vs.\ number of co-located tasks across modalities (DINOv2-Base and Swin-Large).}
    \Description{Mean latency vs.\ number of co-located tasks across modalities (DINOv2-Base and Swin-Large).}
    \label{fig:single_server_models_ntasks}
\end{figure}

\subsubsection{Effect of FM-sharing on Resource Efficiency}

We next show how \systemName scales the number of co-located tasks that a single GPU can serve, and compare against the deployment baselines \BestEffort and \SpatialPartition. \autoref{fig:single_server_ML} reports mean latency as a function of the number of co-located tasks $N$, swept from $N=2$ to $N=10$ on MOMENT-Large at 5 and 7\,RPS per task using Poisson inter-arrival times.
At 5\,RPS per task, \systemName scales smoothly to $N=10$ with mean latency nearly flat, from 28.6 ms at $N=2$ to 60.7 ms at $N=10$.
However, at $N=4$, \SpatialPartition experiences a substantial increase in latency, rising from 34.3 ms at $N=2$ to 114.2 ms at $N=4$. \SpatialPartition collapses as its per-partition compute is exhausted by N=6, where mean latency exceeds 30\,s. While \BestEffort scales smoothly, both \BestEffort and \SpatialPartition run out of memory at $N=10$ when the cumulative cost of 10 backbone replicas exceeds the 16\,GB VRAM budget.
Similarly, at 7\,RPS per task, \systemName's mean latency grows sublinearly from 33\,ms at $N=2$ to 148\,ms at $N=10$, while achieving 79\% lower latency than \BestEffort at $N=8$, the maximum it can run as it reaches the memory limit.
The same scaling behavior holds across modalities (\autoref{fig:single_server_models_ntasks}). On DINOv2-Base (\autoref{fig:single_server_models_ntasks_dino}) and Swin-Large (\autoref{fig:single_server_models_ntasks_swin}), \systemName achieves lower latency up to 90.1\% and 41.7\% improvement over \BestEffort.
These gains of \systemName are because all $N$ tasks share a single backbone and execute as one batch per step, so per-batch GPU compute is amortized across $N$ tasks and total memory sublinearly increases with $N$. \BestEffort replicates the backbone per task, incurring $O(N)$ GPU memory and per-batch compute, which triggers the OOM. \SpatialPartition replicates the backbone and additionally shrinks each task's compute partition as $N$ grows, thereby compounding both costs.

\begin{figure}
    \centering
    \includegraphics[width=0.7\linewidth]{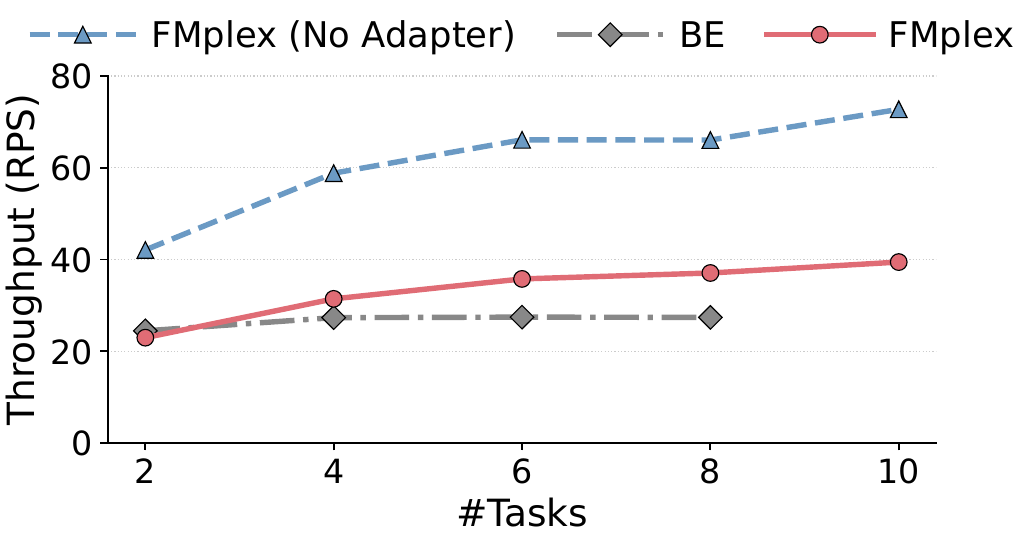}
    \caption{Impact of Customization}
    \Description{Impact of Customization}
    \label{fig:customization}
\end{figure}

\subsubsection{Impact of Customization}

We quantify the throughput cost of per-task customization in \systemName by
comparing it against \BestEffort and a variant of \systemName that serves without LoRA adapters. At $N=2$, \systemName and \BestEffort deliver comparable throughput
(23.0\,RPS vs.\ 24.5\,RPS), since contention is still limited. As $N$ increases, however, \systemName's aggregate throughput grows from 23.0\,RPS at
$N=2$ to 39.5\,RPS at $N=10$ ($1.7\times$), while \BestEffort plateaus near 27\,RPS and fails at $N=10$ due to memory pressure. These gains come from sharing a single base-model forward across tasks. With LoRA adapters, \systemName batches requests over the shared backbone and then executes adapter-specific computation in compatible sub-batches. This improves GPU utilization compared to per-task model invocations in \BestEffort, but it also introduces adapter handling overhead. Relative to
\systemName without adapters, which scales from 42.1\,RPS to 72.7\,RPS over
the same range, per-task customization reduces aggregate throughput by 45\% on average.

\keytakeaway{
\systemName reduces latency by up to 80\% over spatial partitioning and 33.3\% over best-effort co-location for two tasks sharing a backbone, while improving resource efficiency as the number of co-located tasks grows.}

\begin{figure}[t]
    \centering
    \includegraphics[width=1\linewidth]{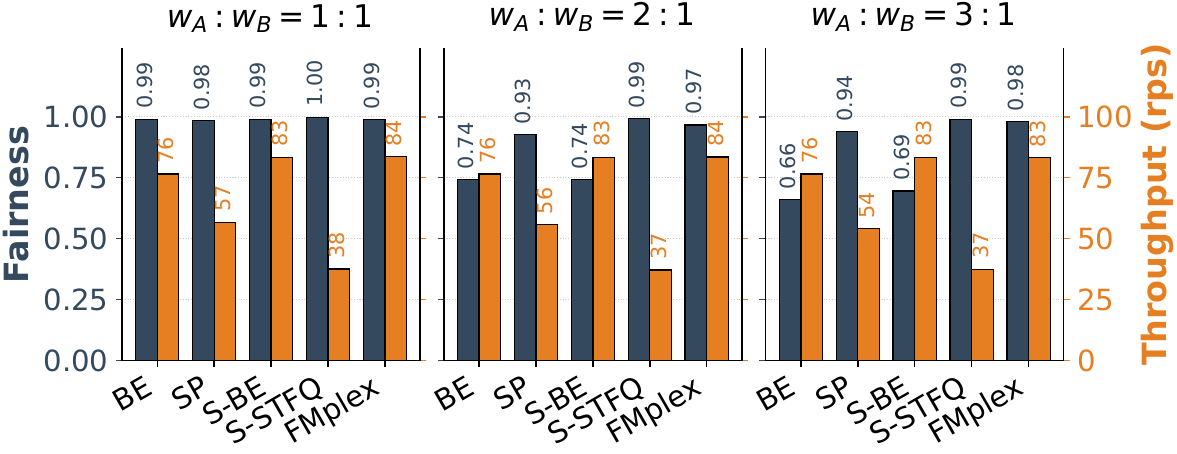}
    \caption{Task-level fairness and aggregate throughput across scheduling approaches for two MOMENT-Large tasks at 60\,RPS each under configured service weights of 1:1, 2:1, and 3:1.}
    \Description{Task-level fairness and aggregate throughput across scheduling approaches for two MOMENT-Large tasks at 60\,RPS each under configured service weights of 1:1, 2:1, and 3:1.}
    \label{fig:isolation_ML}
\end{figure}

\subsection{Fair Sharing and Performance Isolation}\label{sec:eval_isolation}
This section evaluates whether \systemName can preserve task-level service shares and performance isolation when multiple tasks share the same FM backbone. We use \texttt{BFQ} (\autoref{sec:design_isolation}) as the runtime mechanism for enforcing weighted shares while retaining batching efficiency.

\subsubsection{Task-level Fairness under FM-sharing}

To evaluate whether \systemName can provide fair resource allocation without sacrificing system efficiency, we compare \systemName against \BestEffort, \SpatialPartition, \SharedBE, and \SharedSTFQ. \autoref{fig:isolation_ML} reports task-level fairness and aggregate throughput when two MOMENT-Large tasks share one GPU under configured service weights of 1:1, 2:1, and 3:1. In all cases, each client issues requests at 60\,RPS.
In the equal-weight setting (1:1), all methods achieve nearly ideal fairness, with fairness scores between 0.98 and 1.00. \BestEffort is fair in this case because the default GPU scheduler time-slices work from separate CUDA contexts in a round-robin manner~\cite{nvidia-mps}. The main difference is efficiency. \SpatialPartition achieves only 57 \,RPS because each task is confined to a fixed GPU partition, a non-work-conserving approach that significantly affects resource efficiency. Similarly, \SharedSTFQ provides only 38 \,RPS because it disables batching, while still fair, thereby lowering system efficiency.

The asymmetric-weight settings expose the limits of approaches that do not schedule at the task level. With weights of 2:1 and 3:1, \BestEffort and \SharedBE continue to rely on unmanaged GPU sharing, so they do not track the configured 2:1 (resp. 3:1) service weights. Their fairness drops to 0.74 (resp. 0.66) for \BestEffort, and to 0.74 (resp. 0.69) for \SharedBE.
Surprisingly, \SpatialPartition cannot deliver high fairness either. Although it spatially partitions GPU compute, memory contention still affects the underweight task.
Moreover, similar to the equal-weights, \SharedSTFQ achieves near-perfect fairness, with fairness scores of 0.99 in both settings, but its throughput remains limited to 37 \,RPS because it disables batching. In contrast, \systemName preserves fairness at 0.97 and 0.98 while sustaining 84 and 83 \,RPS, demonstrating that task-level scheduling within a shared FM can enforce weighted service shares without compromising batching efficiency.

\begin{figure}
    \centering
    \hfill
    \begin{subfigure}[t]{0.6\linewidth}
        \centering
        \includegraphics[width=\linewidth]{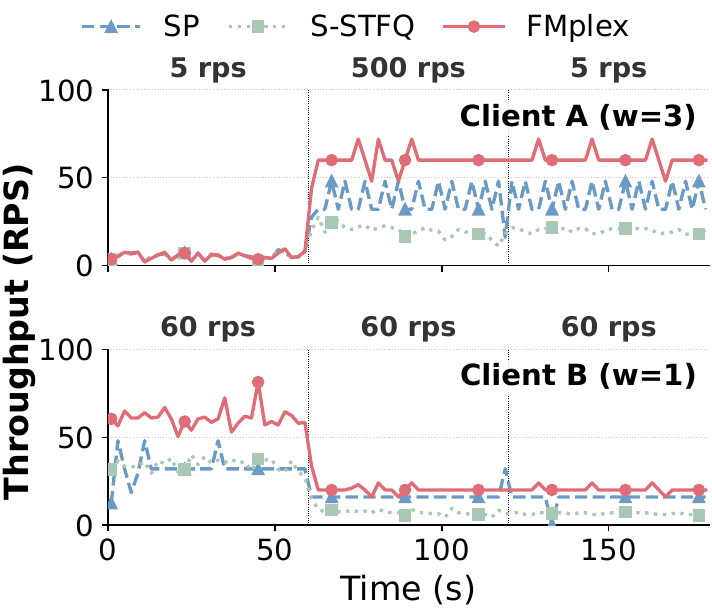}
        \caption{Throughput over time}
        \label{fig:noisy_ML_ts}
    \end{subfigure}
    \hfill
     \begin{subfigure}[t]{0.39\linewidth}
        \centering
        \includegraphics[width=\linewidth]{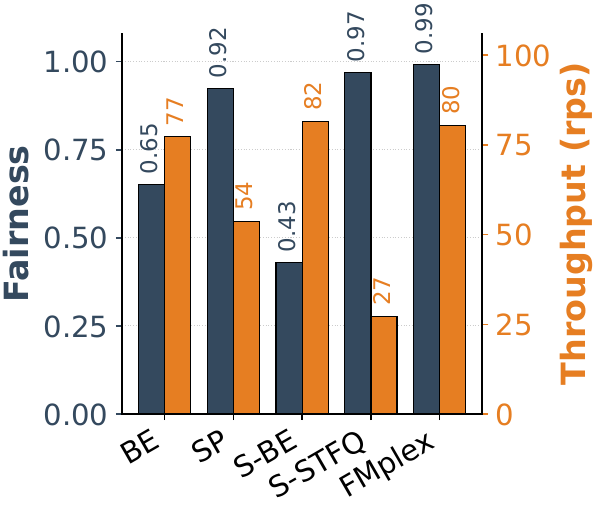}
        \caption{Aggregate results}
        \label{fig:noisy_ML_bar}
    \end{subfigure}
    \hfill
    \hfill
    \caption{Noisy neighbor experiment with weight (3:1) showing throughput and fairness.}
    \Description{Noisy neighbor experiment with weight (3:1) showing throughput and fairness.}
    \label{fig:noisy_ML}
\end{figure}

\subsubsection{Isolation under Noisy-Neighbor Bursts}
Next, we evaluate how \systemName provides performance isolation when one task becomes a noisy neighbor. \autoref{fig:noisy_ML} reports a two-task MOMENT-Large experiment with configured service weights of 3:1. Client B runs at a steady rate of 60\,RPS, while Client A, the high-priority client, starts at 5\,RPS, spikes to 500\,RPS, and then returns to 5\,RPS, a pattern common in serverless and event-driven systems~\cite{Shahrad2020:ServerlessWild}. We compare \systemName against \BestEffort, \SpatialPartition, \SharedBE, and \SharedSTFQ.

\autoref{fig:noisy_ML_ts} shows how each method responds over time to Client A's burst. We omit \BestEffort and \SharedBE for clarity. \SpatialPartition limits Client B's throughput even when Client A's request rate is low, because Client B remains confined to a fixed GPU partition. More importantly, \SpatialPartition does not provide full isolation, where Client B's throughput still drops when Client A increases its demand, as observed in the weighted-sharing experiment.
\SharedSTFQ keeps the two clients closer to their configured service shares, but it sacrifices efficiency because it does not exploit batching. In contrast, \systemName follows the configured 3:1 service allocation during the burst and quickly returns to the steady regime after the burst ends, preserving isolation without discarding batching.
\autoref{fig:noisy_ML_bar} summarizes the same experiment across all approaches. \systemName keeps the tasks close to their configured shares while retaining high throughput, showing that its batch-aware scheduler provides isolation against noisy-neighbor bursts without giving up the efficiency of FM sharing.

\keytakeaway{ \systemName keeps tasks close to their configured shares while retaining high throughput. \systemName sustains 84\,RPS at 0.97--0.98 fairness under asymmetric service weights, while non-batched fair-sharing achieves similar fairness at only 37\,RPS and unmanaged sharing drops fairness to as low as 0.66.}

\subsection{Large-Scale Evaluation}\label{sec:eval_large_scale}
\begin{figure}[t]
    \centering

    \begin{subfigure}{0.27\textwidth}
        \centering
        \includegraphics[width=\linewidth]{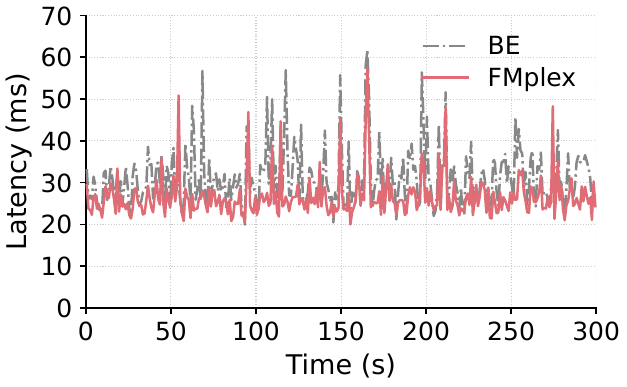}
        \caption{Mean latency over time}
        \label{fig:ls-runtime_mean_time}
    \end{subfigure}
    \hfill
    \begin{subfigure}[b]{0.2\textwidth}
        \centering
        \includegraphics[width=\linewidth]{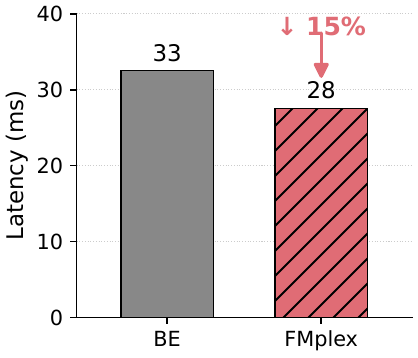}
        \caption{Mean latency}
        \label{fig:ls-runtime_mean_bar}
    \end{subfigure}
    \begin{subfigure}[b]{0.27\textwidth}
        \centering
        \includegraphics[width=\linewidth]{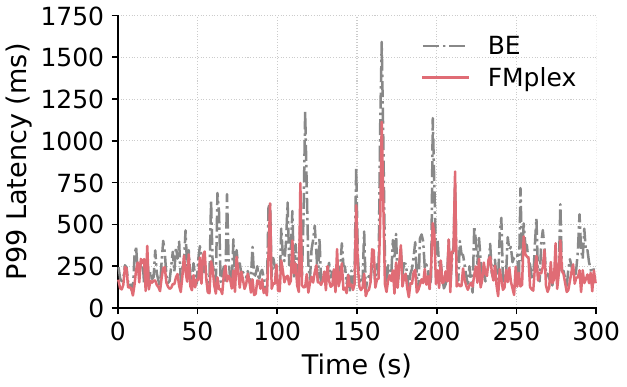}
        \caption{P99 latency over time}
        \label{fig:ls-runtime_99_time}
    \end{subfigure}
    \hfill
    \begin{subfigure}[b]{0.2\textwidth}
        \centering
        \includegraphics[width=\linewidth]{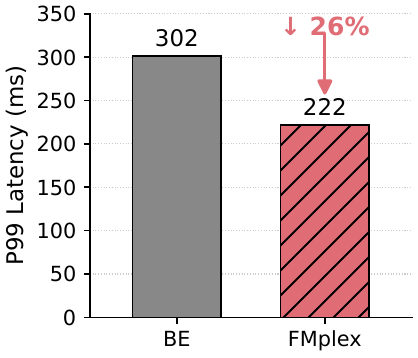}
        \caption{P99 latency}
        \label{fig:ls-runtime_99_bar}
    \end{subfigure}
    \caption{Latency at cluster scale on the 85-task workload: mean (top) and p99 (bottom) over time (left) and across methods (right).}
    \Description{Latency at cluster scale on the 85-task workload: mean (top) and p99 (bottom) over time (left) and across methods (right).}
    \label{fig:ls-runtime}
\end{figure}

\begin{figure}
    \centering
    \vspace{-1em}
    \includegraphics[width=\linewidth]{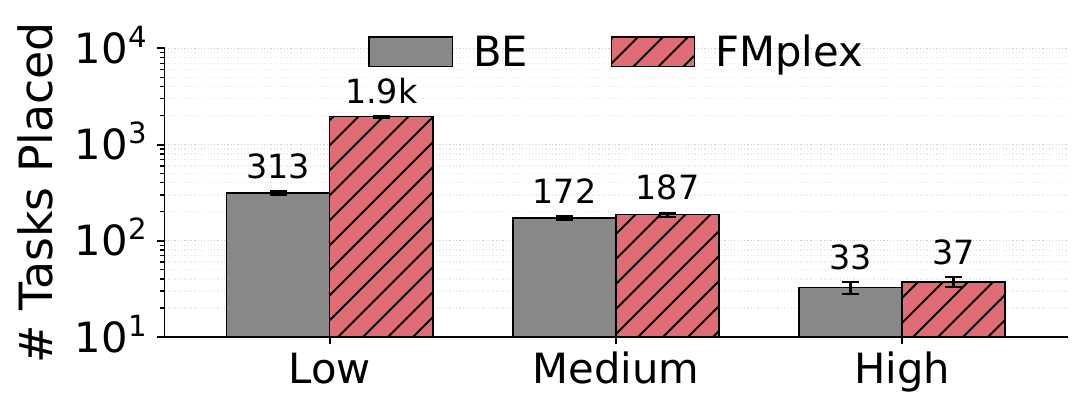}
    \caption{Number of tasks the cluster can host across approaches and load profiles (low, moderate, high).}
    \Description{Number of tasks the cluster can host across approaches and load profiles (low, moderate, high).}
    \label{fig:ls-deployment}
\end{figure}

We now evaluate the \systemName-based serving stack at 16 NVIDIA T4 GPU cluster scale using the Azure Functions trace (\autoref{sec:eval_setup}), with tasks clustered into low-, moderate-, and high-load groups. We compare the stack against \BestEffort along two metrics: end-to-end latency under the resulting shared deployment (\autoref{fig:ls-runtime}) and the number of tasks the cluster can sustain (\autoref{fig:ls-deployment}). We omit \SpatialPartition here since the experiments in \autoref{sec:eval_sharing} already showed it falling behind.

\subsubsection{Cluster-Scale Latency}
\autoref{fig:ls-runtime} reports mean and p99 latency on the 85-task workload across the two approaches, both as a time series over the trace (left column) and aggregated across methods (right column). The \systemName-based stack reduces mean latency by 15\% and p99 latency by 26\% compared to \BestEffort, with the gap widening further during demand bursts. These gains follow from the same mechanisms quantified in \autoref{sec:eval_sharing} and \autoref{sec:eval_isolation}, now operating at cluster scale. The stack forms cross-task batches over each shared backbone and uses \texttt{BFQ} to enforce per-task isolation under bursts, so demand spikes from popular tasks do not propagate to co-located tasks. \BestEffort competes for GPU resources without coordination, missing batching opportunities and inflating tail latency.

\subsubsection{Deployment Efficiency}
\autoref{fig:ls-deployment} reports the number of tasks each approach can sustain across low, moderate, and high load profiles, averaged over 5 runs. The \systemName-based stack hosts up to 6$\times$ more tasks at low load and 8\% and 12\% more at moderate and high load compared to \BestEffort. These gains have two distinct mechanisms across the load regimes. At low load, where memory is the binding constraint, the gain comes from memory amortization, with multiple tasks sharing a single backbone instance rather than each loading its own, raising the admission ceiling. At higher load, where compute becomes the binding constraint, batched execution amortizes per-batch backbone compute across co-resident tasks, so the cluster absorbs more incoming requests before saturating. \BestEffort replicates the backbone per task and forgoes both forms of amortization, capping the number of tasks the cluster can host.

\keytakeaway{At cluster scale, the \systemName-based stack hosts up to 6$\times$ more tasks than \BestEffort at low load and 8--12\% more at moderate and high load by amortizing backbone memory and batched execution across tasks; under the same deployment model, it also reduces mean latency by 15\% and p99 latency by 26\%.}

\subsection{Microbenchmarks}\label{sec:eval_microbenchmarks}

\begin{table}[t]
\centering
\caption{Memory and load time are split into backbone / task-component.
  Latency shows mean per-request backbone and decoder inference time. }
\label{tab:FM_modalities}
\setlength{\tabcolsep}{3pt}
\footnotesize
\begin{tabular}{lrrrrrr}
\toprule
& \multicolumn{2}{c}{\textbf{Memory (MB)}} & \multicolumn{2}{c}{\textbf{Load (ms)}} & \multicolumn{2}{c}{\textbf{Latency (ms)}} \\
\cmidrule(lr){2-3}\cmidrule(lr){4-5}\cmidrule(lr){6-7}
\textbf{Backbone} & BB & Task & BB & Task & BB & Task \\
\midrule
\textbf{Time Series} \\
~~Moment-Large    & 1462      & 0.52 & 5737 &   24.98 &  23.63 & 0.66 \\
~Papageip~   &  23.24 & 0.26 & 161.69 & 5.05 & 15.79 & 0.33 \\
\midrule
\textbf{Vision} \\
~~DINOv2-Base    & 347      & 0.03 &    817  &   0.17 &  18.86 & 0.37 \\
~~Swin-Large    & 347      & 0.04 &    1001  &   0.18 &  30.89 & 0.35 \\
\midrule
\textbf{LLM} \\
~~Qwen2.5-3B   & 6285  & - & 3095 & -   &  310.19  & -  \\
~~Mistral-7B   & 14496  & - & 5927 & -   & 604.34 & -  \\
\midrule
\textbf{VLM} \\
~~Qwen2-VL-2B   & 4420  & 8.76 & 4492 & 176.53   &    134.51  & -  \\
\bottomrule
\end{tabular}
\end{table}
\subsubsection{Backbone and Task Costs}
\autoref{tab:FM_modalities} reports per-backbone memory, load time, and inference latency on \systemName, decomposed into the backbone (BB) and task-specific component (Task), across time-series, vision, LLM, and VLM modalities. For backbones with task-specific components, the BB component dominates memory by 1--3 orders of magnitude (e.g., 1462\,MB vs 0.52\,MB on MOMENT-Large) and load time by similar factors (5737\,ms vs 25\,ms). This asymmetry is what \systemName's deployment sharing amortizes (\autoref{sec:eval_sharing}), since once a backbone is loaded, additional tasks add only the lightweight task component. Inference latency follows the same pattern, with BB execution dominating per-request cost across modalities.

\subsubsection{Workload Adaptation}\label{sec:eval_adaptation}
\begin{figure}[t]
    \centering

        \includegraphics[width=\linewidth]        {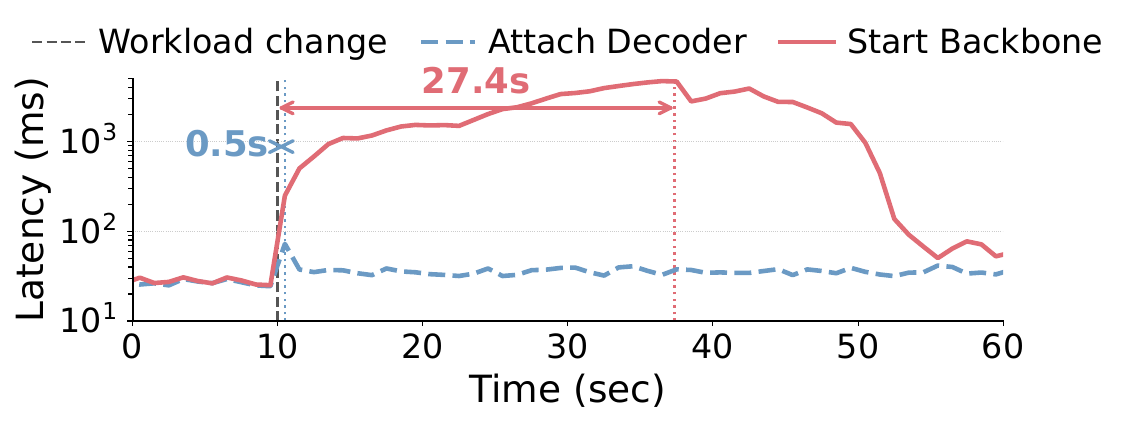}
     \caption{Adaptation latency after a workload surge in \systemName and \BestEffort.}
    \Description{Adaptation latency after a workload surge in \systemName and \BestEffort.}
    \label{fig:adaptation_ML}
\end{figure}

A key benefit of decoupling tasks from FM backbones is that \systemName can adapt by moving task-specific state instead of reloading the full FM, as described in \autoref{sec:orch_elastic}. \autoref{fig:adaptation_ML} demonstrates this benefit through the behavior of \systemName and \BestEffort for a MOMENT-Large task after a workload surge. For \systemName, a MOMENT-Large instance is already available on a different server, so \systemEngine replicates the task by attaching its decoder to that resident backbone and distributing load across both replicas. This path completes in 500\,ms and produces only a transient increase in latency.
In \BestEffort, there is no backbone sharing, so the system must start a new MOMENT-Large instance before it can shift load\footnote{While this can also happen for \systemName, this represents the worst-case scenario.}.  This start-backbone path waits until the new backbone is ready, around 58\,s after the workload change. During this interval, mean latency rises by roughly two orders of magnitude, and the backlog continues to affect latency even after the backbone becomes available. The result highlights why \systemName's \vFM abstraction matters for adaptation. When a compatible physical FM already exists, adaptation operates at the task-state timescale rather than the backbone-loading timescale.

\begin{figure}
    \vspace{-1em}
    \centering
    \includegraphics[width=\linewidth]{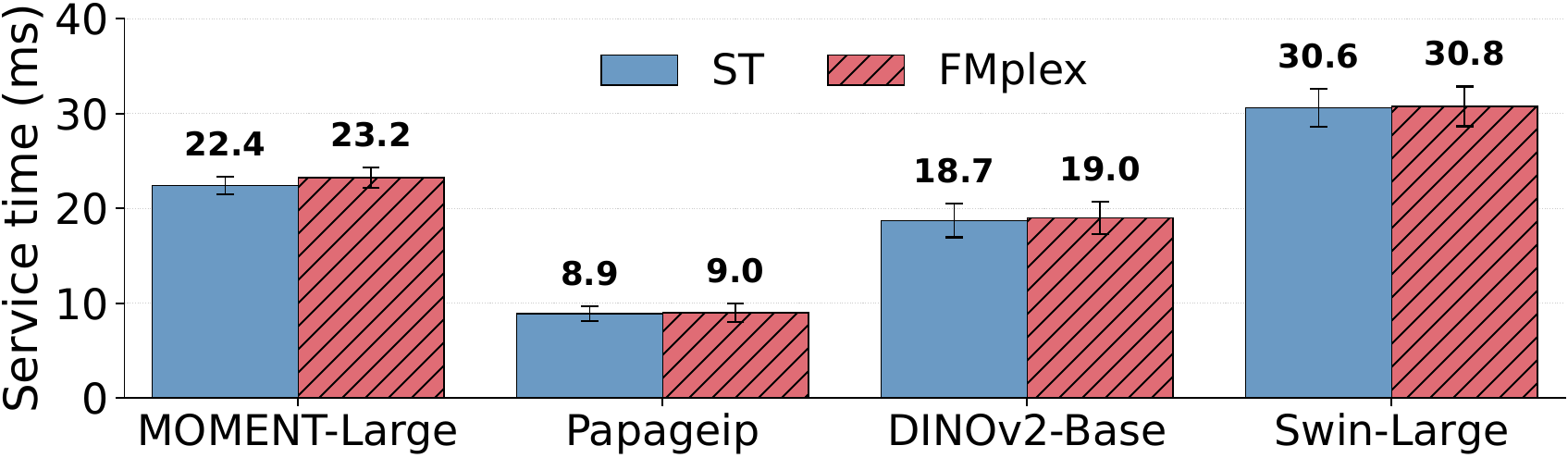}
    \caption{\systemName scheduling overhead.}
    \Description{\systemName scheduling overhead.}
    \label{fig:latency_overhead}
\end{figure}

\subsubsection{Overheads}
\autoref{fig:latency_overhead} reports per-request service time across backbones at batch-size 1, comparing execution under \systemName against direct backbone execution without the \systemName scheduler. The \systemName overhead is below 0.8\,ms in all cases and 0.35\,ms on average, remaining small relative to the backbone forward pass across all modalities. This overhead is dominated by per-task queue handling and \texttt{BFQ}'s tag computation, both of which are independent of batch size and small compared to backbone execution time.

\section{Related Work}
\label{sec:related}

\noindent\textbf{\textit{Model Serving and Sharing.}}
Model-serving systems~\cite{Daniel2017:Clipper, Gujarati2020:Clockwork, Romero2021:INFaaS, Ahmad2024:Proteus, nvidia_triton, Nigenda2022:SageMaker, aws_bedrock} treat each model as the atomic unit of deployment. Systems such as Nexus~\cite{Shen2019:Nexus} and AlpaServe~\cite{Li2023:AlpaServe} reuse engines or replicas across requests but not backbones across tasks.
Adapter-sharing systems such as S-LoRA~\cite{Sheng2024:S-LoRA}, Punica~\cite{Chen2024:Punica}, dLoRA~\cite{Wu2024:dLoRA} and EdgeLoRA~\cite{Shen2025:EdgeLoRA} share a LLM base across many LoRA adapters and provide highly optimized kernels for this case. These systems looks at the space with one backbone, one extension type and one modality. \systemName generalizes this point to an abstraction layer, the \vFM, that admits aribitary extension type (decoders, encoders, adapters), heterogeneous backbones and independent task lifecycles. For LLM-with-LoRA workloads, \systemName can use adapter-sharing systems as its execution backend while the vFM abstraction operates one layer above it.

\noindent\textbf{\textit{Fairness and Isolation in LLM Serving.}}
Fair scheduling has been studied recently for LLM serving, with VTC~\cite{Sheng2024:LLMFair} and related work~\cite{ Shen2024:FastSwitch, Cao2025:Locality-LLM-Fair} proposing schedulers that provide weighted fair scheduling across clients of a shared LLM. \systemName's scheduling component \texttt{BFQ}, is weighted fair-queueing variant applied at the \vFM boundary to arbitrate among tasks sharing a backbone, Within an LLM-backed vFM, token-level schedulers such as VTC can be composed with \texttt{BFQ}.

\noindent\textbf{\textit{Virtualization and Resource Management.}}
The \vFM abstraction extends the classical virtualization pattern of exposing logically private views of shared physical resources~\cite{Popek1974:VMRequirements, Creasy1981:VM370, Merkel2014:Docker,Smith2005:VMs,lxc} to foundation models. A \vFM is a task-scoped view of the backbone, with isolation enforced through per-task queues, accounting, and scheduling rather than replication. Classical virtualization platforms also recognized that co-located VMs share memory pages from common OS images, libraries, and applications and sharing-aware colocation algorithms can amplify this benefit through smarter placement~\cite{Sindelar2011:VMColocation,Wood2009:Memorybuddies}. The setting of \systemName takes this one step further, the shared substrate is the FM bbackbone itself, which dominates the memory and execution cost, and \systemEngine's sharing-aware deployment policy is the FM-serving analog of sharing-aware VM colocation.

\section{Conclusion}
We presented \systemName, a virtualization layer for foundation-model serving that decouples downstream tasks from the physical deployment of FM backbones. Through the \vFM abstraction, \systemName lets independently customized tasks share a physical FM instance while preserving task-specific state, independent lifecycles, and task-level performance isolation.
A supporting runtime scheduler and lightweight sharing-aware controller make this deployment model practical across heterogeneous FMs and workloads. Across 16 foundation-model backbones and 92 downstream tasks, \systemName reduces latency by up to 80\% over spatial partitioning and 33.3\% over best-effort co-location, while hosting up to 6$\times$ more tasks at cluster scale. Our future work will explore richer placement and resource-management policies on top of the FM virtualization abstraction.

\begin{acks} 
This research is supported by National Science Foundation (NSF) grants 2213636, 2211302, 2211888, 2325956, 2213636, 2105494, 2211301, and US Army grant W911NF-17-2-0196.
\end{acks}

\bibliographystyle{ACM-Reference-Format}
\bibliography{papers, DNNbibs}

@inproceedings{Redmon2016:Yolo,
  title={{You Only Look Once: Unified, Real-Time Object Detection}},
  author={Redmon, Joseph and Divvala, Santosh and Girshick, Ross and Farhadi, Ali},
  booktitle={{Proceedings of the IEEE Conference on Computer Vision and Pattern Recognition (CVPR)}},
  pages={779--788},
  year={2016}
}

@inproceedings{efficient-net,
  title     = {{EfficientNet: Rethinking Model Scaling for Convolutional Neural Networks}},
  author    = {Tan, Mingxing and Le, Quoc},
  booktitle = {Proceedings of the 36th International Conference on Machine Learning},
  pages     = {6105--6114},
  year      = {2019},
  volume    = {97},
  series    = {Proceedings of Machine Learning Research},
  month     = {09--15 Jun},
  publisher = {PMLR},
  url       = {https://proceedings.mlr.press/v97/tan19a.html}
}

@inproceedings{resnet,
  title     = {Deep {R}esidual {L}earning for {I}mage {R}ecognition},
  author    = {He, Kaiming and Zhang, Xiangyu and Ren, Shaoqing and Sun, Jian},
  booktitle = {Proceedings of the IEEE conference on computer vision and pattern recognition (CVPR)},
  year      = {2016}
}

@inproceedings{Goswami2024:MOMENT,
  title={{MOMENT: A Family of Open Time-series Foundation Models}},
  author={Mononito Goswami and Konrad Szafer and Arjun Choudhry and Yifu Cai and Shuo Li and Artur Dubrawski},
  booktitle={International Conference on Machine Learning},
  year={2024}
}

@article{Ansari2024:Chronos,
  title={Chronos: Learning the Language of Time Series},
  author={Ansari, Abdul Fatir and
          Stella, Lorenzo and
          Turkmen, Caner and
          Zhang, Xiyuan and Mercado, Pedro and
          Shen, Huibin and
          Shchur, Oleksandr and
          Rangapuram, Syama Syndar and
          Pineda Arango, Sebastian and
          Kapoor, Shubham and
          Zschiegner, Jasper and
          Maddix, Danielle C. and
          Mahoney, Michael W. and
          Torkkola, Kari and
          Gordon Wilson, Andrew and
          Bohlke-Schneider, Michael and
          Wang, Yuyang},
  journal={Transactions on Machine Learning Research},
  issn={2835-8856},
  year={2024},
  url={https://openreview.net/forum?id=gerNCVqqtR}
}

@article{
Oquab2024:DinoV2,
title={{DINOv2: Learning Robust Visual Features without Supervision}},
author={Maxime Oquab and Timoth{\'e}e Darcet and Th{\'e}o Moutakanni and Huy V. Vo and Marc Szafraniec and Vasil Khalidov and Pierre Fernandez and Daniel HAZIZA and Francisco Massa and Alaaeldin El-Nouby and Mido Assran and Nicolas Ballas and Wojciech Galuba and Russell Howes and Po-Yao Huang and Shang-Wen Li and Ishan Misra and Michael Rabbat and Vasu Sharma and Gabriel Synnaeve and Hu Xu and Herve Jegou and Julien Mairal and Patrick Labatut and Armand Joulin and Piotr Bojanowski},
journal={Transactions on Machine Learning Research},
issn={2835-8856},
year={2024},
}

@article{Hu2022:LoRA,
  title={{LoRA: Low-Rank Adaptation of Large Language Models}},
  author={Hu, Edward J and Shen, Yelong and Wallis, Phillip and Allen-Zhu, Zeyuan and Li, Yuanzhi and Wang, Shean and Wang, Lu and Chen, Weizhu and others},
  journal={ICLR},
  volume={1},
  number={2},
  pages={3},
  year={2022}
}

@inbook{Shen2025:EdgeLoRA,
author = {Shen, Zheyu and He, Yexiao and Wang, Ziyao and Zhang, Yuning and Sun, Guoheng and Ye, Wanghao and Li, Ang},
title = {{EdgeLoRA: An Efficient Multi-Tenant LLM Serving System on Edge Devices}},
year = {2025},
isbn = {9798400714535},
publisher = {Association for Computing Machinery},
address = {New York, NY, USA},
url = {https://doi.org/10.1145/3711875.3729141},
booktitle = {Proceedings of the 23rd Annual International Conference on Mobile Systems, Applications and Services},
pages = {138–153},
numpages = {16}
}

@misc{Sheng2024:S-LoRA,
      title={{S-LoRA: Serving Thousands of Concurrent LoRA Adapters}}, 
      author={Ying Sheng and Shiyi Cao and Dacheng Li and Coleman Hooper and Nicholas Lee and Shuo Yang and Christopher Chou and Banghua Zhu and Lianmin Zheng and Kurt Keutzer and Joseph E. Gonzalez and Ion Stoica},
      year={2024},
      eprint={2311.03285},
      archivePrefix={arXiv},
      primaryClass={cs.LG},
      url={https://arxiv.org/abs/2311.03285}, 
}

@misc{nvidia_triton,
  author       = {{NVIDIA}},
  title        = {{Triton Inference Server}},
  year         = {2024},
  url          = {https://developer.nvidia.com/triton-inference-server},
  note         = {Accessed: 2025-04-13}
}

@INPROCEEDINGS{Liang2023:ETF,
  author={Liang, Qianlin and Hanafy, Walid A. and Bashir, Noman and Irwin, David and Shenoy, Prashant},
  booktitle={{2023 IEEE/ACM Symposium on Edge Computing (SEC)}}, 
  title={{Energy Time Fairness: Balancing Fair Allocation of Energy and Time for GPU Workloads}}, 
  year={2023},
  volume={},
  number={},
  pages={53-66},
  doi={10.1145/3583740.3628435}
}

@inproceedings{Satya2021:TheRole,
author = {Satyanarayanan, Mahadev and Beckmann, Nathan and Lewis, Grace A. and Lucia, Brandon},
title = {{The Role of Edge Offload for Hardware-Accelerated Mobile Devices}},
year = {2021},
isbn = {9781450383233},
url = {https://doi.org/10.1145/3446382.3448360},
doi = {10.1145/3446382.3448360},
booktitle = {Proceedings of the 22nd International Workshop on Mobile Computing Systems and Applications},
pages = {22–29},
numpages = {8},
location = {Virtual, United Kingdom},
series = {HotMobile '21}
}

@inproceedings {Gujarati2020:Clockwork,
    author = {Arpan Gujarati and Reza Karimi and Safya Alzayat and Wei Hao and Antoine Kaufmann and Ymir Vigfusson and Jonathan Mace},
    title = {{Serving DNNs like Clockwork: Performance Predictability from the Bottom Up}},
    booktitle = {14th USENIX Symposium on Operating Systems Design and Implementation (OSDI 20)},
    year = {2020},
    isbn = {978-1-939133-19-9},
    pages = {443--462},
    url = {https://www.usenix.org/conference/osdi20/presentation/gujarati},
    publisher = {USENIX Association},
    month = nov,
}

@inproceedings {Daniel2017:Clipper,
author = {Daniel Crankshaw and Xin Wang and Guilio Zhou and Michael J. Franklin and Joseph E. Gonzalez and Ion Stoica},
title = {{Clipper: A Low-Latency Online Prediction Serving System}},
booktitle = {14th USENIX Symposium on Networked Systems Design and Implementation (NSDI 17)},
year = {2017},
isbn = {978-1-931971-37-9},
address = {Boston, MA},
pages = {613--627},
url = {https://www.usenix.org/conference/nsdi17/technical-sessions/presentation/crankshaw},
publisher = {USENIX Association},
month = mar,
}

@inproceedings {Romero2021:INFaaS,
author = {Francisco Romero and Qian Li and Neeraja J. Yadwadkar and Christos Kozyrakis},
title = {{INFaaS}: Automated Model-less Inference Serving},
booktitle = {2021 USENIX Annual Technical Conference (USENIX ATC 21)},
year = {2021},
isbn = {978-1-939133-23-6},
pages = {397--411},
url = {https://www.usenix.org/conference/atc21/presentation/romero},
publisher = {USENIX Association},
month = jul
}

@inproceedings {Yu2022:Orca,
author = {Gyeong-In Yu and Joo Seong Jeong and Geon-Woo Kim and Soojeong Kim and Byung-Gon Chun},
title = {{Orca: A Distributed Serving System for {Transformer-Based} Generative Models}},
booktitle = {16th USENIX Symposium on Operating Systems Design and Implementation (OSDI 22)},
year = {2022},
isbn = {978-1-939133-28-1},
address = {Carlsbad, CA},
pages = {521--538},
url = {https://www.usenix.org/conference/osdi22/presentation/yu},
publisher = {USENIX Association},
month = jul
}

@inproceedings{Nigenda2022:SageMaker,
author = {Nigenda, David and Karnin, Zohar and Zafar, Muhammad Bilal and Ramesha, Raghu and Tan, Alan and Donini, Michele and Kenthapadi, Krishnaram},
title = {{Amazon SageMaker Model Monitor: A System for Real-Time Insights into Deployed Machine Learning Models}},
year = {2022},
isbn = {9781450393850},
publisher = {Association for Computing Machinery},
address = {New York, NY, USA},
url = {https://doi.org/10.1145/3534678.3539145},
doi = {10.1145/3534678.3539145},
booktitle = {Proceedings of the 28th ACM SIGKDD Conference on Knowledge Discovery and Data Mining},
pages = {3671–3681},
numpages = {11},
location = {Washington DC, USA},
series = {KDD '22}
}

@inproceedings{Shen2019:Nexus,
author = {Shen, Haichen and Chen, Lequn and Jin, Yuchen and Zhao, Liangyu and Kong, Bingyu and Philipose, Matthai and Krishnamurthy, Arvind and Sundaram, Ravi},
title = {{Nexus: A GPU Cluster Engine for Accelerating DNN-Based Video Analysis}},
year = {2019},
isbn = {9781450368735},
publisher = {Association for Computing Machinery},
address = {New York, NY, USA},
url = {https://doi.org/10.1145/3341301.3359658},
doi = {10.1145/3341301.3359658},
booktitle = {Proceedings of the 27th ACM Symposium on Operating Systems Principles},
pages = {322–337},
numpages = {16},
location = {Huntsville, Ontario, Canada},
series = {SOSP '19}
}

@ARTICLE{Zhang202:MArk,  
    author={Zhang, Chengliang and Yu, Minchen and wang, wei and Yan, Feng},  
    journal={IEEE Transactions on Cloud Computing},  title={{Enabling Cost-Effective, SLO-Aware Machine Learning Inference Serving on Public Cloud}}, 
    year={2020}, 
    volume={}, 
    number={},  
    pages={1-1},  
    doi={10.1109/TCC.2020.3006751}
}

@article{Liang2020:Queuing,
    author = {Liang, Qianlin and Hanafy, Walid A. and Ali-Eldin, Ahmed and Shenoy, Prashant},
    title = {{Model-Driven Cluster Resource Management for AI Workloads in Edge Clouds}},
    year = {2023},
    issue_date = {March 2023},
    volume = {18},
    number = {1},
    issn = {1556-4665},
    url = {https://doi.org/10.1145/3582080},
    doi = {10.1145/3582080},
    journal = {ACM Transactions on Autonomous and Adaptive Systems},
    month = {mar},
    articleno = {2},
    numpages = {26}
}

@article{Siam2025:AIoT,
author = {Siam, Shakhrul Iman and Ahn, Hyunho and Liu, Li and Alam, Samiul and Shen, Hui and Cao, Zhichao and Shroff, Ness and Krishnamachari, Bhaskar and Srivastava, Mani and Zhang, Mi},
title = {{Artificial Intelligence of Things: A Survey}},
year = {2025},
issue_date = {January 2025},
publisher = {Association for Computing Machinery},
address = {New York, NY, USA},
volume = {21},
number = {1},
issn = {1550-4859},
url = {https://doi.org/10.1145/3690639},
doi = {10.1145/3690639},
journal = {ACM Trans. Sen. Netw.},
month = jan,
articleno = {9},
numpages = {75}
}

@article{Bommasani2021:FoundationModels,
title={{On the Opportunities and Risks of Foundation Models}},
author={Rishi Bommasani and others},
journal={ArXiv},
year={2021},
url={https://crfm.stanford.edu/assets/report.pdf}
}

@inproceedings{Ahmad2024:Proteus,
author = {Ahmad, Sohaib and Guan, Hui and Friedman, Brian D. and Williams, Thomas and Sitaraman, Ramesh K. and Woo, Thomas},
title = {{Proteus: A High-Throughput Inference-Serving System with Accuracy Scaling}},
year = {2024},
isbn = {9798400703720},
url = {https://doi.org/10.1145/3617232.3624849},
doi = {10.1145/3617232.3624849},
booktitle = {Proceedings of the 29th ACM International Conference on Architectural Support for Programming Languages and Operating Systems, Volume 1},
pages = {318–334},
numpages = {17},
location = {La Jolla, CA, USA},
series = {ASPLOS '24}
}

@inproceedings{Sindelar2011:VMColocation,
author = {Sindelar, Michael and Sitaraman, Ramesh K. and Shenoy, Prashant},
title = {Sharing-aware algorithms for virtual machine colocation},
year = {2011},
isbn = {9781450307437},
publisher = {Association for Computing Machinery},
address = {New York, NY, USA},
url = {https://doi.org/10.1145/1989493.1989554},
doi = {10.1145/1989493.1989554},
pages = {367–378},
numpages = {12},
location = {San Jose, California, USA},
series = {SPAA '11}
}

@article{Wood2009:Memorybuddies,
author = {Wood, Timothy and Tarasuk-Levin, Gabriel and Shenoy, Prashant and Desnoyers, Peter and Cecchet, Emmanuel and Corner, Mark D.},
title = {Memory buddies: exploiting page sharing for smart colocation in virtualized data centers},
year = {2009},
issue_date = {July 2009},
publisher = {Association for Computing Machinery},
address = {New York, NY, USA},
volume = {43},
number = {3},
issn = {0163-5980},
url = {https://doi.org/10.1145/1618525.1618529},
doi = {10.1145/1618525.1618529},
journal = {SIGOPS Oper. Syst. Rev.},
month = jul,
pages = {27–36},
numpages = {10}
}

@article{Smith2005:VMs,
  author   = {Smith, J.E. and Ravi Nair},
  journal  = {Computer},
  title    = {{The architecture of Virtual Machines}},
  year     = {2005},
  volume   = {38},
  number   = {5},
  pages    = {32-38},
  doi      = {10.1109/MC.2005.173}
}

@manual{cuda_green_contexts,
  title        = {CUDA Driver API: Green Contexts},
  author       = {{NVIDIA Corporation}},
  year         = {2026},
  url          = {https://docs.nvidia.com/cuda/cuda-driver-api/group__CUDA__GREEN__CONTEXTS.html},
  note         = {Accessed: 2026-03-27},
  organization = {NVIDIA Corporation}
}

@misc{nvidia_mig,
  title  = {{NVIDIA Multi-Instance GPU (MIG)}},
  author = {{NVIDIA Corporation}},
  year   = {2026},
  url    = {https://www.nvidia.com/en-us/technologies/multi-instance-gpu/},
  note   = {Accessed: 2026-03-27}
}

@inproceedings{Houlsby2019:PETL,
  title      = {Parameter-Efficient Transfer Learning for {NLP}},
  author     = {Houlsby, Neil and Giurgiu, Andrei and Jastrzebski, Stanislaw and Morrone, Bruna and De Laroussilhe, Quentin and Gesmundo, Andrea and Attariyan, Mona and Gelly, Sylvain},
  booktitle  = {Proceedings of the 36th International Conference on Machine Learning},
  pages      = {2790--2799},
  year       = {2019},
  editor     = {Chaudhuri, Kamalika and Salakhutdinov, Ruslan},
  volume     = {97},
  series     = {Proceedings of Machine Learning Research},
  month      = {09--15 Jun},
  _publisher = {PMLR}
}

@misc{Xu2023:PEFTLLM,
  title         = {{Parameter-Efficient Fine-Tuning Methods for Pretrained Language Models: A Critical Review and Assessment}},
  author        = {Lingling Xu and Haoran Xie and Si-Zhao Joe Qin and Xiaohui Tao and Fu Lee Wang},
  year          = {2023},
  eprint        = {2312.12148},
  archiveprefix = {arXiv},
  primaryclass  = {cs.CL},
  url           = {https://arxiv.org/abs/2312.12148}
}

@misc{Balne2024:PEFTAnalysis,
  title         = {{Parameter Efficient Fine Tuning: A Comprehensive Analysis Across Applications}},
  author        = {Charith Chandra Sai Balne and Sreyoshi Bhaduri and Tamoghna Roy and Vinija Jain and Aman Chadha},
  year          = {2024},
  eprint        = {2404.13506},
  archiveprefix = {arXiv},
  primaryclass  = {cs.LG},
  url           = {https://arxiv.org/abs/2404.13506}
}

@misc{Rao2025:LLMFinance,
  title         = {LLMs Meet Finance: Fine-Tuning Foundation Models for the Open FinLLM Leaderboard},
  author        = {Varun Rao and Youran Sun and Mahendra Kumar and Tejas Mutneja and Agastya Mukherjee and Haizhao Yang},
  year          = {2025},
  eprint        = {2504.13125},
  archiveprefix = {arXiv},
  primaryclass  = {cs.CL},
  url           = {https://arxiv.org/abs/2504.13125}
}

@misc{Dong2024:SEGMedical,
  title         = {Segment anything model 2: an application to 2D and 3D medical images},
  author        = {Haoyu Dong and Hanxue Gu and Yaqian Chen and Jichen Yang and Yuwen Chen and Maciej A. Mazurowski},
  year          = {2024},
  eprint        = {2408.00756},
  archiveprefix = {arXiv},
  primaryclass  = {cs.CV},
  url           = {https://arxiv.org/abs/2408.00756}
}

@inproceedings{LSTM,
  title     = {Long short-term memory recurrent neural network architectures for large scale acoustic modeling},
  author    = {Hasim Sak and Andrew W. Senior and Françoise Beaufays},
  year      = {2014},
  booktitle = {INTERSPEECH},
  pages     = {338-342}
}

@misc{Brown2020:FewShot,
  title         = {{Language Models are Few-Shot Learners}},
  author        = {Tom B. Brown and Benjamin Mann and Nick Ryder and Melanie Subbiah and Jared Kaplan and Prafulla Dhariwal and Arvind Neelakantan and Pranav Shyam and Girish Sastry and Amanda Askell and Sandhini Agarwal and Ariel Herbert-Voss and Gretchen Krueger and Tom Henighan and Rewon Child and Aditya Ramesh and Daniel M. Ziegler and Jeffrey Wu and Clemens Winter and Christopher Hesse and Mark Chen and Eric Sigler and Mateusz Litwin and Scott Gray and Benjamin Chess and Jack Clark and Christopher Berner and Sam McCandlish and Alec Radford and Ilya Sutskever and Dario Amodei},
  year          = {2020},
  eprint        = {2005.14165},
  archiveprefix = {arXiv},
  primaryclass  = {cs.CL},
  url           = {https://arxiv.org/abs/2005.14165}
}

@article{GEMMA2024,
  title   = {{Gemma: Open Models Based on Gemini Research and Technology}},
  author  = {Mesnard, Thomas and Hardin, Cassidy and Dadashi, Robert and Bhupatiraju, Surya and Pathak, Shreya and Sifre, Laurent and Rivi{\`e}re, Morgane and Kale, Mihir Sanjay and Love, Juliette and others},
  journal = {arXiv preprint arXiv:2403.08295},
  year    = {2024}
}

@article{Guo2025:DeepSeek,
  title      = {{DeepSeek-R1 incentivizes reasoning in LLMs through Reinforcement Learning}},
  volume     = {645},
  issn       = {1476-4687},
  url        = {http://dx.doi.org/10.1038/s41586-025-09422-z},
  doi        = {10.1038/s41586-025-09422-z},
  number     = {8081},
  journal    = {Nature},
  _publisher = {Springer Science and Business Media LLC},
  author     = {Guo, Daya and Yang, Dejian and others},
  year       = {2025},
  month      = sep,
  pages      = {633–638}
}

@misc{touvron2023:LLaMA,
  title         = {{LLaMA: Open and Efficient Foundation Language Models}},
  author        = {Hugo Touvron and Thibaut Lavril and Gautier Izacard and Xavier Martinet and Marie-Anne Lachaux and Timothée Lacroix and Baptiste Rozière and Naman Goyal and Eric Hambro and Faisal Azhar and Aurelien Rodriguez and Armand Joulin and Edouard Grave and Guillaume Lample},
  year          = {2023},
  eprint        = {2302.13971},
  archiveprefix = {arXiv},
  primaryclass  = {cs.CL},
  url           = {https://arxiv.org/abs/2302.13971}
}

@inproceedings{Yang2016:IQA,
  author    = {Yang, Zichao and He, Xiaodong and Gao, Jianfeng and Deng, Li and Smola, Alex},
  title     = {Stacked Attention Networks for Image Question Answering},
  booktitle = {Proceedings of the IEEE Conference on Computer Vision and Pattern Recognition (CVPR)},
  month     = {June},
  year      = {2016}
}

@misc{Huang2025:MedTS,
  title         = {Repurposing Foundation Model for Generalizable Medical Time Series Classification},
  author        = {Nan Huang and Haishuai Wang and Zihuai He and Marinka Zitnik and Xiang Zhang},
  year          = {2025},
  eprint        = {2410.03794},
  archiveprefix = {arXiv},
  primaryclass  = {cs.LG},
  url           = {https://arxiv.org/abs/2410.03794}
}

@misc{pillai2025:PaPaGei,
  title         = {PaPaGei: Open Foundation Models for Optical Physiological Signals},
  author        = {Arvind Pillai and Dimitris Spathis and Fahim Kawsar and Mohammad Malekzadeh},
  year          = {2025},
  eprint        = {2410.20542},
  archiveprefix = {arXiv},
  primaryclass  = {cs.LG},
  url           = {https://arxiv.org/abs/2410.20542}
}

@article{feofanov2025:Mantis,
  title   = {{Mantis: Lightweight Calibrated Foundation Model for User-Friendly Time Series Classification}},
  author  = {Vasilii Feofanov and Songkang Wen and Marius Alonso and Romain Ilbert and Hongbo Guo and Malik Tiomoko and Lujia Pan and Jianfeng Zhang and Ievgen Redko},
  journal = {arXiv preprint arXiv:2502.15637},
  year    = {2025}
}

@misc{zhu2025:FinCast,
  title         = {{FinCast: A Foundation Model for Financial Time-Series Forecasting}},
  author        = {Zhuohang Zhu and Haodong Chen and Qiang Qu and Vera Chung},
  year          = {2025},
  eprint        = {2508.19609},
  archiveprefix = {arXiv},
  primaryclass  = {cs.LG},
  url           = {https://arxiv.org/abs/2508.19609}
}

@misc{shi2025:kronos,
  title         = {{Kronos: A Foundation Model for the Language of Financial Markets}},
  author        = {Yu Shi and Zongliang Fu and Shuo Chen and Bohan Zhao and Wei Xu and Changshui Zhang and Jian Li},
  year          = {2025},
  eprint        = {2508.02739},
  archiveprefix = {arXiv},
  primaryclass  = {q-fin.ST},
  url           = {https://arxiv.org/abs/2508.02739}
}

@misc{Gnassounou2025:TSEEG,
  title         = {{Leveraging Generic Time Series Foundation Models for EEG Classification}},
  author        = {Théo Gnassounou and Yessin Moakher and Shifeng Xie and Vasilii Feofanov and Ievgen Redko},
  year          = {2025},
  eprint        = {2510.27522},
  archiveprefix = {arXiv},
  primaryclass  = {cs.LG},
  url           = {https://arxiv.org/abs/2510.27522}
}

@inproceedings{Kumar2025:MixForecast,
  author     = {Kumar, Anuj and Saravanan, Harish Kumar and Dwivedi, Shivam and Arjunan, Pandarasamy},
  title      = {MixForecast: Mixer-Enhanced Foundation Model for Load Forecasting},
  year       = {2025},
  isbn       = {9798400716089},
  _publisher = {Association for Computing Machinery},
  _address   = {New York, NY, USA},
  url        = {https://doi.org/10.1145/3722565.3727193},
  doi        = {10.1145/3722565.3727193},
  booktitle  = {Proceedings of the 2nd International Workshop on Foundation Models for Cyber-Physical Systems \& Internet of Things},
  pages      = {25–30},
  numpages   = {6},
  keywords   = {Demand-side Load management, Energy Forecasting, Energy Informatics, Machine Learning, Short-term Load Forecasting (STLF), Smart Grid, Time Series Foundation Models (TSFM)},
  location   = {Irvine, CA, USA},
  series     = {FMSys}
}

@misc{Simeone2026:TSFM-Energy,
  title         = {{Time Series Foundation Models for Energy Load Forecasting on Consumer Hardware: A Multi-Dimensional Zero-Shot Benchmark}},
  author        = {Luigi Simeone},
  year          = {2026},
  eprint        = {2602.10848},
  archiveprefix = {arXiv},
  primaryclass  = {cs.LG},
  url           = {https://arxiv.org/abs/2602.10848}
}

@misc{Maji2025:CarbonX,
  title         = {{CarbonX: An Open-Source Tool for Computational Decarbonization Using Time Series Foundation Models}},
  author        = {Diptyaroop Maji and Kang Yang and Prashant Shenoy and Ramesh K Sitaraman and Mani Srivastava},
  year          = {2025},
  eprint        = {2510.01521},
  archiveprefix = {arXiv},
  primaryclass  = {cs.LG},
  url           = {https://arxiv.org/abs/2510.01521}
}

@inproceedings{Dosovitskiy2021:ViT,
  title     = {{An Image is Worth 16x16 Words: Transformers for Image Recognition at Scale}},
  author    = {Dosovitskiy, Alexey and Beyer, Lucas and Kolesnikov, Alexander and Weissenborn, Dirk and Zhai, Xiaohua and Unterthiner, Thomas and Dehghani, Mostafa and Minderer, Matthias and Heigold, Georg and Gelly, Sylvain and Uszkoreit, Jakob and Houlsby, Neil},
  booktitle = {International Conference on Learning Representations (ICLR)},
  year      = {2021},
  url       = {https://openreview.net/forum?id=YicbFdNTTy}
}

@misc{grpc-framework,
  title        = {gRPC: A high performance open-source universal {RPC} framework},
  howpublished = {\url{https://grpc.io/}},
  author       = {{gRPC Authors}},
  year         = {2026},
  note         = {Accessed: 2026-03-26}
}

@misc{peft,
  title        = {{PEFT: State-of-the-art Parameter-Efficient Fine-Tuning methods}},
  author       = {Sourab Mangrulkar and Sylvain Gugger and Lysandre Debut and Younes Belkada and Sayak Paul and Benjamin Bossan and Marian Tietz},
  howpublished = {\url{https://github.com/huggingface/peft}},
  year         = {2022}
}

@manual{nvidia-mps,
  title        = {NVIDIA Multi-Process Service},
  author       = {{NVIDIA Corporation}},
  organization = {NVIDIA Corporation},
  year         = {2026},
  url          = {https://docs.nvidia.com/deploy/mps/index.html}
}

@inbook{Paszke2019:Pytorch,
  author     = {Paszke, Adam and Gross, Sam and Massa, Francisco and Lerer, Adam and Bradbury, James and Chanan, Gregory and Killeen, Trevor and Lin, Zeming and Gimelshein, Natalia and Antiga, Luca and Desmaison, Alban and K\"{o}pf, Andreas and Yang, Edward and DeVito, Zach and Raison, Martin and Tejani, Alykhan and Chilamkurthy, Sasank and Steiner, Benoit and Fang, Lu and Bai, Junjie and Chintala, Soumith},
  title      = {PyTorch: an imperative style, high-performance deep learning library},
  year       = {2019},
  _publisher = {Curran Associates Inc.},
  _address   = {Red Hook, NY, USA},
  booktitle  = {Proceedings of the 33rd International Conference on Neural Information Processing Systems},
  articleno  = {721},
  numpages   = {12}
}

@inproceedings{Kwon2023:vLLM,
  title     = {{Efficient Memory Management for Large Language Model Serving with PagedAttention}},
  author    = {Woosuk Kwon and Zhuohan Li and Siyuan Zhuang and Ying Sheng and Lianmin Zheng and Cody Hao Yu and Joseph E. Gonzalez and Hao Zhang and Ion Stoica},
  booktitle = {Proceedings of the ACM SIGOPS 29th Symposium on Operating Systems Principles},
  year      = {2023}
}

@inproceedings{bakita2023:TPCs,
  title     = {{Hardware Compute Partitioning on {NVIDIA} {GPUs}}},
  author    = {Bakita, Joshua and Anderson, James H},
  booktitle = {Proceedings of the 29th IEEE Real-Time and Embedded Technology and Applications Symposium},
  year      = {2023},
  month     = {May},
  pages     = {54--66},
  _series   = {RTAS}
}

@inproceedings{Radford2021:CLIP,
  title      = {{Learning Transferable Visual Models From Natural Language Supervision}},
  author     = {Radford, Alec and Kim, Jong Wook and Hallacy, Chris and Ramesh, Aditya and Goh, Gabriel and Agarwal, Sandhini and Sastry, Girish and Askell, Amanda and Mishkin, Pamela and Clark, Jack and Krueger, Gretchen and Sutskever, Ilya},
  booktitle  = {Proceedings of the 38th International Conference on Machine Learning},
  pages      = {8748--8763},
  year       = {2021},
  editor     = {Meila, Marina and Zhang, Tong},
  volume     = {139},
  series     = {Proceedings of Machine Learning Research},
  month      = {18--24 Jul},
  _publisher = {PMLR},
  pdf        = {http://proceedings.mlr.press/v139/radford21a/radford21a.pdf},
  url        = {https://proceedings.mlr.press/v139/radford21a.html}
}

@misc{Baris2025:FMs-CPS-IoT,
  title         = {{Foundation Models for CPS-IoT: Opportunities and Challenges}},
  author        = {Ozan Baris and Yizhuo Chen and Gaofeng Dong and Liying Han and Tomoyoshi Kimura and Pengrui Quan and Ruijie Wang and Tianchen Wang and Tarek Abdelzaher and Mario Bergés and Paul Pu Liang and Mani Srivastava},
  year          = {2025},
  eprint        = {2501.16368},
  archiveprefix = {arXiv},
  primaryclass  = {cs.LG},
  url           = {https://arxiv.org/abs/2501.16368}
}

@misc{Meta2024:Llama32-Vision,
  title        = {Llama 3.2 Vision models},
  author       = {{Meta}},
  year         = {2024},
  howpublished = {\url{https://www.llama.com/docs/model-cards-and-prompt-formats/llama3_2/}},
  note         = {Accessed: 2026-03-05}
}

@misc{Yao2024:MiniCPMV,
  title         = {{MiniCPM-V: A GPT-4V Level MLLM on Your Phone}},
  author        = {Yuan Yao and Tianyu Yu and Ao Zhang and Chongyi Wang and Junbo Cui and Hongji Zhu and Tianchi Cai and Haoyu Li and Weilin Zhao and Zhihui He and Qianyu Chen and Huarong Zhou and Zhensheng Zou and Haoye Zhang and Shengding Hu and Zhi Zheng and Jie Zhou and Jie Cai and Xu Han and Guoyang Zeng and Dahai Li and Zhiyuan Liu and Maosong Sun},
  year          = {2024},
  eprint        = {2408.01800},
  archiveprefix = {arXiv},
  primaryclass  = {cs.CV},
  url           = {https://arxiv.org/abs/2408.01800}
}

@misc{Wang2024:Qwen2VL,
  title         = {{Qwen2-VL: Enhancing Vision-Language Model's Perception of the World at Any Resolution}},
  author        = {Peng Wang and Shuai Bai and Sinan Tan and Shijie Wang and Zhihao Fan and Jinze Bai and Keqin Chen and Xuejing Liu and Jialin Wang and Wenbin Ge and Yang Fan and Kai Dang and Mengfei Du and Xuancheng Ren and Rui Men and Dayiheng Liu and Chang Zhou and Jingren Zhou and Junyang Lin},
  year          = {2024},
  eprint        = {2409.12191},
  archiveprefix = {arXiv},
  primaryclass  = {cs.CV},
  url           = {https://arxiv.org/abs/2409.12191}
}

@inproceedings{Liu2023:LLaVA,
  author     = {Liu, Haotian and Li, Chunyuan and Wu, Qingyang and Lee, Yong Jae},
  title      = {Visual instruction tuning},
  year       = {2023},
  _publisher = {Curran Associates Inc.},
  _address   = {Red Hook, NY, USA},
  booktitle  = {Proceedings of the 37th International Conference on Neural Information Processing Systems},
  articleno  = {1516},
  numpages   = {25},
  location   = {New Orleans, LA, USA},
  series     = {NIPS '23}
}

@inproceedings{Sheng2024:LLMFair,
  author     = {Ying Sheng and Shiyi Cao and Dacheng Li and Banghua Zhu and Zhuohan Li and Danyang Zhuo and Joseph E. Gonzalez and Ion Stoica},
  title      = {{Fairness in Serving Large Language Models}},
  booktitle  = {18th USENIX Symposium on Operating Systems Design and Implementation (OSDI 24)},
  year       = {2024},
  isbn       = {978-1-939133-40-3},
  _address   = {Santa Clara, CA},
  pages      = {965--988},
  url        = {https://www.usenix.org/conference/osdi24/presentation/sheng},
  _publisher = {USENIX Association},
  month      = jul
}

@misc{Shen2024:FastSwitch,
  title         = {{FastSwitch: Optimizing Context Switching Efficiency in Fairness-aware Large Language Model Serving}},
  author        = {Ao Shen and Zhiyao Li and Mingyu Gao},
  year          = {2024},
  eprint        = {2411.18424},
  archiveprefix = {arXiv},
  primaryclass  = {cs.LG},
  url           = {https://arxiv.org/abs/2411.18424}
}

@misc{Cao2025:Locality-LLM-Fair,
  title         = {{Locality-aware Fair Scheduling in LLM Serving}},
  author        = {Shiyi Cao and Yichuan Wang and Ziming Mao and Pin-Lun Hsu and Liangsheng Yin and Tian Xia and Dacheng Li and Shu Liu and Yineng Zhang and Yang Zhou and Ying Sheng and Joseph Gonzalez and Ion Stoica},
  year          = {2025},
  eprint        = {2501.14312},
  archiveprefix = {arXiv},
  primaryclass  = {cs.DC},
  url           = {https://arxiv.org/abs/2501.14312}
}

@inproceedings{Li2023:AlpaServe,
  author     = {Li, Zhuohan and Zheng, Lianmin and Zhong, Yinmin and Liu, Vincent and Sheng, Ying and Jin, Xin and Huang, Yanping and Chen, Zhifeng and Zhang, Hao and Gonzalez, Joseph E. and Stoica, Ion},
  title      = {{AlpaServe}: Statistical Multiplexing with Model Parallelism for Deep Learning Serving},
  booktitle  = {17th USENIX Symposium on Operating Systems Design and Implementation (OSDI 23)},
  year       = {2023},
  pages      = {663--679},
  _publisher = {USENIX Association}
}

@inproceedings{Chen2024:Punica,
  author    = {Chen, Lequn and Ye, Zihao and Wu, Yongji and Zhuo, Danyang and Ceze, Luis and Krishnamurthy, Arvind},
  title     = {Punica: Multi-Tenant {LoRA} Serving},
  booktitle = {Proceedings of Machine Learning and Systems (MLSys)},
  year      = {2024}
}

@inproceedings{Wu2024:dLoRA,
  author     = {Wu, Bingyang and Zhu, Ruidong and Zhang, Zili and Sun, Peng and Liu, Xuanzhe and Jin, Xin},
  title      = {{dLoRA}: Dynamically Orchestrating Requests and Adapters for {LoRA} {LLM} Serving},
  booktitle  = {18th USENIX Symposium on Operating Systems Design and Implementation (OSDI 24)},
  year       = {2024},
  _publisher = {USENIX Association}
}

@article{Popek1974:VMRequirements,
  author  = {Popek, Gerald J. and Goldberg, Robert P.},
  title   = {Formal Requirements for Virtualizable Third Generation Architectures},
  journal = {Communications of the ACM},
  volume  = {17},
  number  = {7},
  pages   = {412--421},
  year    = {1974},
  doi     = {10.1145/361011.361073}
}

@article{Creasy1981:VM370,
  author  = {Creasy, R. J.},
  title   = {The Origin of the {VM/370} Time-Sharing System},
  journal = {IBM Journal of Research and Development},
  volume  = {25},
  number  = {5},
  pages   = {483--490},
  year    = {1981},
  doi     = {10.1147/rd.255.0483}
}

@article{Merkel2014:Docker,
  author  = {Merkel, Dirk},
  title   = {{Docker: Lightweight Linux Containers for Consistent Development and Deployment}},
  journal = {Linux Journal},
  volume  = {2014},
  number  = {239},
  year    = {2014}
}

@misc{lxc,
  author       = {{LXC Contributors}},
  title        = {{LXC: Linux Containers}},
  year         = {2026},
  howpublished = {\url{https://linuxcontainers.org/}},
  note         = {Accessed: 2026-04-14}
}

@article{WFQ,
  author     = {Demers, A. and Keshav, S. and Shenker, S.},
  title      = {{Analysis and Simulation of a Fair Queueing Algorithm}},
  year       = {1989},
  issue_date = {Sep. 1989},
  _publisher = {Association for Computing Machinery},
  _address   = {New York, NY, USA},
  volume     = {19},
  number     = {4},
  issn       = {0146-4833},
  url        = {https://doi.org/10.1145/75247.75248},
  doi        = {10.1145/75247.75248},
  journal    = {SIGCOMM Comput. Commun. Rev.},
  month      = {aug},
  pages      = {1–12},
  numpages   = {12}
}

@article{STFQ,
  author     = {Goyal, Pawan and Vin, Harrick M. and Cheng, Haichen},
  title      = {{Start-Time Fair Queueing: A Scheduling Algorithm for Integrated Services Packet Switching Networks}},
  year       = {1997},
  issue_date = {Oct. 1997},
  _publisher = {IEEE Press},
  volume     = {5},
  number     = {5},
  issn       = {1063-6692},
  url        = {https://doi.org/10.1109/90.649569},
  doi        = {10.1109/90.649569},
  journal    = {IEEE/ACM Trans. Netw.},
  month      = {oct},
  pages      = {690–704},
  numpages   = {15},
  keywords   = {packet scheduling, integrated services networks, fair queueing}
}

@article{GPS,
  author  = {Parekh, A.K. and Gallager, R.G.},
  journal = {IEEE/ACM Transactions on Networking},
  title   = {{A generalized processor sharing approach to flow control in integrated services networks: the single-node case}},
  year    = {1993},
  volume  = {1},
  number  = {3},
  pages   = {344-357},
  doi     = {10.1109/90.234856}
}

@inproceedings{waldspurger1994lottery,
  title     = {{Lottery scheduling: Flexible Proportional-share Resource Management}},
  author    = {Waldspurger, Carl A and Weihl, William E},
  booktitle = {Proceedings of the 1st USENIX conference on Operating Systems Design and Implementation},
  pages     = {1--es},
  year      = {1994}
}

@inproceedings{ng2024TailClipper2,
  _address   = {New York, NY, USA},
  series     = {{SoCC} '24},
  title      = {{TailClipper}: {Reducing} {Tail} {Response} {Time} of {Distributed} {Services} {Through} {System}-{Wide} {Scheduling}},
  isbn       = {979-8-4007-1286-9},
  shorttitle = {{TailClipper}},
  url        = {https://doi.org/10.1145/3698038.3698554},
  doi        = {10.1145/3698038.3698554},
  urldate    = {2025-12-31},
  booktitle  = {Proceedings of the 2024 {ACM} {Symposium} on {Cloud} {Computing}},
  _publisher = {Association for Computing Machinery},
  author     = {Ng, Nathan and Souza, Abel and Ali-Eldin, Ahmed and Irwin, David and Towsley, Don and Shenoy, Prashant},
  month      = nov,
  year       = {2024},
  pages      = {398--414}
}

@inproceedings{Shahrad2020:ServerlessWild,
  author    = {Mohammad Shahrad and Rodrigo Fonseca and Inigo Goiri and Gohar Chaudhry and Paul Batum and Jason Cooke and Eduardo Laureano and Colby Tresness and Mark Russinovich and Ricardo Bianchini},
  title     = {{Serverless in the Wild: Characterizing and Optimizing the Serverless Workload at a Large Cloud Provider}},
  booktitle = {2020 USENIX Annual Technical Conference (USENIX ATC 20)},
  year      = {2020},
  isbn      = {978-1-939133-14-4},
  pages     = {205--218},
  url       = {https://www.usenix.org/conference/atc20/presentation/shahrad},
  publisher = {USENIX Association},
  month     = jul
}

@inproceedings{Standley2020:MultiTask,
  author    = {Standley, Trevor and Zamir, Amir and Chen, Dawn and Guibas, Leonidas and Malik, Jitendra and Savarese, Silvio},
  title     = {Which tasks should be learned together in multi-task learning?},
  year      = {2020},
  publisher = {JMLR.org},
  booktitle = {Proceedings of the 37th International Conference on Machine Learning},
  articleno = {846},
  numpages  = {13},
  series    = {ICML'20}
}

@inproceedings{Liu2019:MultiTask,
  author    = {Liu, Shikun and Johns, Edward and Davison, Andrew J.},
  title     = {End-To-End Multi-Task Learning With Attention},
  booktitle = {Proceedings of the IEEE/CVF Conference on Computer Vision and Pattern Recognition (CVPR)},
  month     = {June},
  year      = {2019}
}

@misc{aws_bedrock,
  author       = {{Amazon Web Services}},
  title        = {Amazon Bedrock},
  year         = {2026},
  howpublished = {\url{https://aws.amazon.com/bedrock/}},
  note         = {Accessed: 2026-05-14}
}

@misc{liu2021:SwinTransformer,
      title={{Swin Transformer: Hierarchical Vision Transformer using Shifted Windows}}, 
      author={Ze Liu and Yutong Lin and Yue Cao and Han Hu and Yixuan Wei and Zheng Zhang and Stephen Lin and Baining Guo},
      year={2021},
      eprint={2103.14030},
      archivePrefix={arXiv},
      primaryClass={cs.CV},
      url={https://arxiv.org/abs/2103.14030}, 
}

@misc{jiang2023:mistral7b,
      title={{Mistral 7B}}, 
      author={Albert Q. Jiang and Alexandre Sablayrolles and Arthur Mensch and Chris Bamford and Devendra Singh Chaplot and Diego de las Casas and Florian Bressand and Gianna Lengyel and Guillaume Lample and Lucile Saulnier and Lélio Renard Lavaud and Marie-Anne Lachaux and Pierre Stock and Teven Le Scao and Thibaut Lavril and Thomas Wang and Timothée Lacroix and William El Sayed},
      year={2023},
      eprint={2310.06825},
      archivePrefix={arXiv},
      primaryClass={cs.CL},
      url={https://arxiv.org/abs/2310.06825}, 
}

@INPROCEEDINGS{fairness_metric,
  author={Elliott, R.},
  booktitle={IEEE CCECE2002. Canadian Conference on Electrical and Computer Engineering. Conference Proceedings (Cat. No.02CH37373)}, 
  title={A measure of fairness of service for scheduling algorithms in multiuser systems}, 
  year={2002},
  volume={3},
  number={},
  pages={1583-1588 vol.3},
  doi={10.1109/CCECE.2002.1012991}}
\newpage

\appendix
\onecolumn
\section{\FMAPI Training Script}\label{app:full_listign}

Listing~\ref{lst:moment_full} demonstrates a full training example using \systemName. The example constructs a task pipeline, fine-tuning it through \FMAPI, and running inference. It uses the MOMENT-Base backbone with a custom encoder, an MLP decoder, and a LoRA adapter for heart rate prediction on the PPG-BP dataset.

\section{Benefits of FM-sharing}\label{app:fm_sharing}

\begin{figure*}[ht!]
    \centering
    \includegraphics[width=0.9\linewidth]{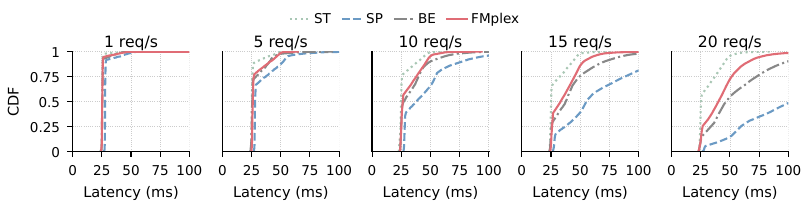}
    \caption{CDF across request rates for MOMENT-Large (\autoref{fig:two_tasks_ML})}
    \Description{CDF across request rates for MOMENT-Large (\autoref{fig:two_tasks_ML})}
    \label{fig:two_tasks_ML_CDF}
\end{figure*}

\begin{figure*}[ht!]
    \centering
    \includegraphics[width=0.9\linewidth]{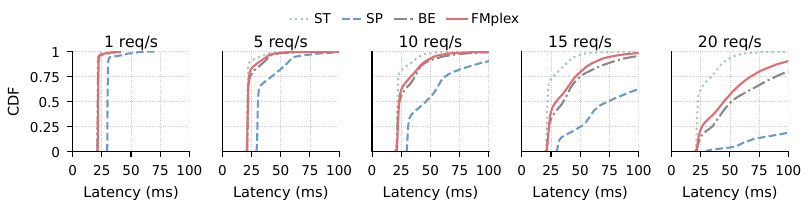}
    \caption{CDF across request rates for DINOv2-Base (\autoref{fig:two_tasks_models_dino})}
    \Description{CDF across request rates for DINOv2-Base (\autoref{fig:two_tasks_models_dino})}
    \label{fig:two_tasks_dinobase_CDF}
\end{figure*}

\begin{figure*}[ht!]
    \centering
    \includegraphics[width=0.9\linewidth]{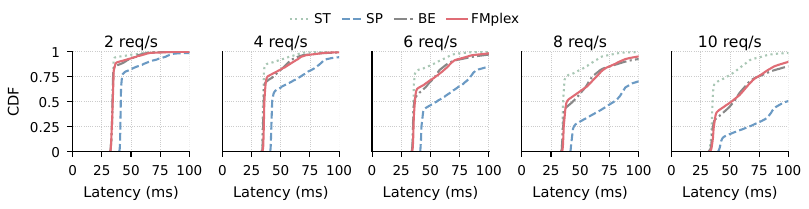}
    \caption{CDF across request rates for Swin-Large (\autoref{fig:two_tasks_models_swin})}
    \Description{CDF across request rates for Swin-Large (\autoref{fig:two_tasks_models_swin})}
    \label{fig:two_tasks_swinlarge_CDF}
\end{figure*}

This appendix complements the latency results in \autoref{sec:eval_sharing} with full latency CDFs across request rates for the two-task experiments. We report CDFs for MOMENT-Large (\autoref{fig:two_tasks_ML_CDF}), DINOv2-Base (\autoref{fig:two_tasks_dinobase_CDF}), and Swin-Large (\autoref{fig:two_tasks_swinlarge_CDF}). Across all three backbones and request rates, \systemName's latency distribution dominates \BestEffort and \SpatialPartition in both the body and the tail.

\begin{listing*}[ht!]
\centering
\begin{minipage}{\linewidth}
\begin{minted}[fontsize=\small, linenos, numbersep=8pt, xleftmargin=1.6em, breaklines=true]{python}
lora_config = LoraConfig(r=64, lora_alpha=32, target_modules=["q", "v"], lora_dropout=0.05)
P=Pipeline(vMomentModelBase(model))
P.add_encoder(LinearChannelCombiner(cfg={num_channels=3,new_num_channels=1}, load=True))
P.add_decoder(MLPDecoder(cfg={'input_dim':1024,'output_dim':1,'hidden_dim':128},load=True))
P.attach_adapter(lora_config)
P.train(dataloader_train,parts_to_train=['encoder','decoder','adapter'],cfg=task_cfg['train_config'])
y_test,y_pred=P.run(dataloader_test,cfg=task_cfg['inference_config'])
\end{minted}
\end{minipage}\hfill
\vspace{1em}
\centering\caption{Example of a complete \FMAPI training and inference pipeline attaching the Moment backbone with an MLP decoder, a linear-channel encoder, and a LoRA adapter for heart rate prediction on the PPG-BP dataset.}
\label{lst:moment_full}
\end{listing*}

\end{document}